\documentclass[11pt]{article}

\usepackage[margin=1in]{geometry}
\usepackage[T1]{fontenc}
\usepackage[utf8]{inputenc}
\DeclareUnicodeCharacter{2011}{\mbox{-}} 
\DeclareUnicodeCharacter{2010}{-}       
\usepackage{amsmath,amssymb}
\usepackage{graphicx} 
\usepackage{color}
\usepackage[numbers,sort&compress]{natbib}
\usepackage[hidelinks]{hyperref}

\begin{document}

\title{Evolving view on phylogenetic networks}

\author{Claudia Sol\'is-Lemus$^{1,\ast}$\\[0.5em]
\parbox{0.9\textwidth}{\centering\small
$^1$ Wisconsin Institute for Discovery, Department of Plant Pathology, University of Wisconsin-Madison, 500 Lincoln Dr, Madison, WI 53706\\[0.3em]
$^{\ast}$ Correspondence to be sent to: University of Wisconsin-Madison, 500 Lincoln Dr, Madison, WI 53706, USA. \\Email:~solislemus@wisc.edu}}
\date{}

\maketitle

\begin{abstract}
Reticulate processes such as hybridization, introgression, and horizontal gene transfer cannot be fully represented by a bifurcating tree. Enter phylogenetic networks: first as split graphs to visualize tree discordance, then as explicit probabilistic models that capture biological phenomena. 
Here, we describe the broad taxonomy of network representations, distinguishing the principal classes of explicit networks, their biological interpretability and our ability to accurately estimate them from empirical data.  
We also trace the evolution of the main network inferential methods from hybrid detection tests, distance- and subgraph-based amalgamation methods, probabilistic approaches under the multispecies network coalescent, composite-likelihood and divide-and-conquer frameworks, while highlighting the selective pressures of statistical identifiability and computational scalability that have shaped this evolution. 
As we move towards a network thinking paradigm, previously isolated methodological lineages from population genetics, phylogenomics, and mathematical network theory are now introgressing, uniting diverse network models into a shared framework that can integrate sequence‑ and species‑level reticulate processes, increase robustness to systematic errors, and refine algorithms for genome‑scale data, expanding the tree of life into a richer, more entangled yet clearer picture of evolution.
\end{abstract}


\section{Shifting into the era of network thinking}

Science is driven forward by paradigm shifts \citep{Kuhn1962}: 
moments when
scientists recognize that anomalies are not just margin points, but
aspects from reality ignored by our current
paradigm. The acknowledgment of anomalies is a difficult task because
it involves thinking outside the current scientific model, but it is
only through such a change in perspective that knowledge can expand and
evolve.

The tree paradigm fails to explain certain patterns (anomalies under the current paradigm) just like any scientific paradigm. 
Gene flow
is such an anomaly. It represents a real evolutionary process that violates the key assumption of tree thinking: 
vertical inheritance of genetic material. Historically, gene flow in
the form of horizontal gene transfer or hybridization 
has often been set aside to simplify 
the reconstruction of the evolutionary histories of
species. Hybrid species or genes with strong conflicting histories
tend to be removed from analysis as noise that does not fit the
overall tree-like pattern. 
Gene flow, however, is not rare. One could argue it is even the norm in some taxonomic groups.
Gene flow appears at all levels in the tree of life as shown
by a large body of evidence \citep[e.g. in recent years]{morales2018evidence, macguigan2019phylogenomic, karimi2020reticulate, olave2020implementing, morales2021disentangling, mcguire2023species, folk2023identifying, sanderson2023phylogenomics, myers2024phylogenomic, pyron2025complex},
among which we can highlight pervasive hybridization during rapid plant radiations (e.g., \textit{Tillandsia} \citep{yardeni2025explosive}) and extensive introgression across both deep and shallow phylogenetic timescales in animal systems such as \textit{Heliconius} butterflies \citep{thawornwattana2022full}.

%
%

The discovery of such reticulate evolutionary histories is only possible through the development of hybrid detection and network inference methods, as traditional tree-based methods are unable to capture these processes.
Recently, gene flow has begun to appear
as an integral part of phylogenomic analyses, effectively shifting the paradigm from tree thinking to network thinking. 
%

The era of network thinking brings several challenges that will be described in this perspective. Even the concept of a phylogenetic network is not straightforward, as different classes of networks reflect distinct biological interpretations.
Furthermore, estimating phylogenetic networks is computationally more demanding than inferring trees and necessitates larger quantities of high-quality data. 
In particular, statistical identifiability and computational scalability remain major challenges due to the complexity of models that must simultaneously account for multiple biological processes like incomplete lineage sorting (ILS) and gene flow.
Furthermore, optimization within the expansive network space remains operationally difficult.

Despite these challenges, 
the continued advancement of methodological approaches is increasingly allowing evolutionary biologists to infer reticulate events across diverse lineages.
The more we untangle the truth about evolution and species diversity, the more complicated it appears, but it is only
because we are seeing it closely. The reality is more complex than we originally envisioned. So, as we zoom in, pieces will certainly not fit in the way we initially imagined. As we expand our understanding and inference capabilities of phylogenetic networks,  we are able to include more reticulate organisms in the tree -- or rather network -- of life, effectively providing a broader and more complete picture of the evolutionary history of biodiversity on Earth.


\section{A web of definitions for networks}
\label{def}

Mathematically, trees are structurally simple. Trees are directed acyclic graphs composed of external nodes (or leaves) which represent extant taxa and internal nodes that represent speciation events of ancestral populations \citep{semple2003phylogenetics}. Edges connecting nodes represent ancestry, and along the edges, genetic mutations are expected to accumulate. Thus, a tree is a mathematical abstraction used to represent the evolutionary process (sequence of speciation events) by which an ancestral population gives rise to a collection of extant taxa.

\begin{figure}
    \centering
    \includegraphics[width=0.4\linewidth]{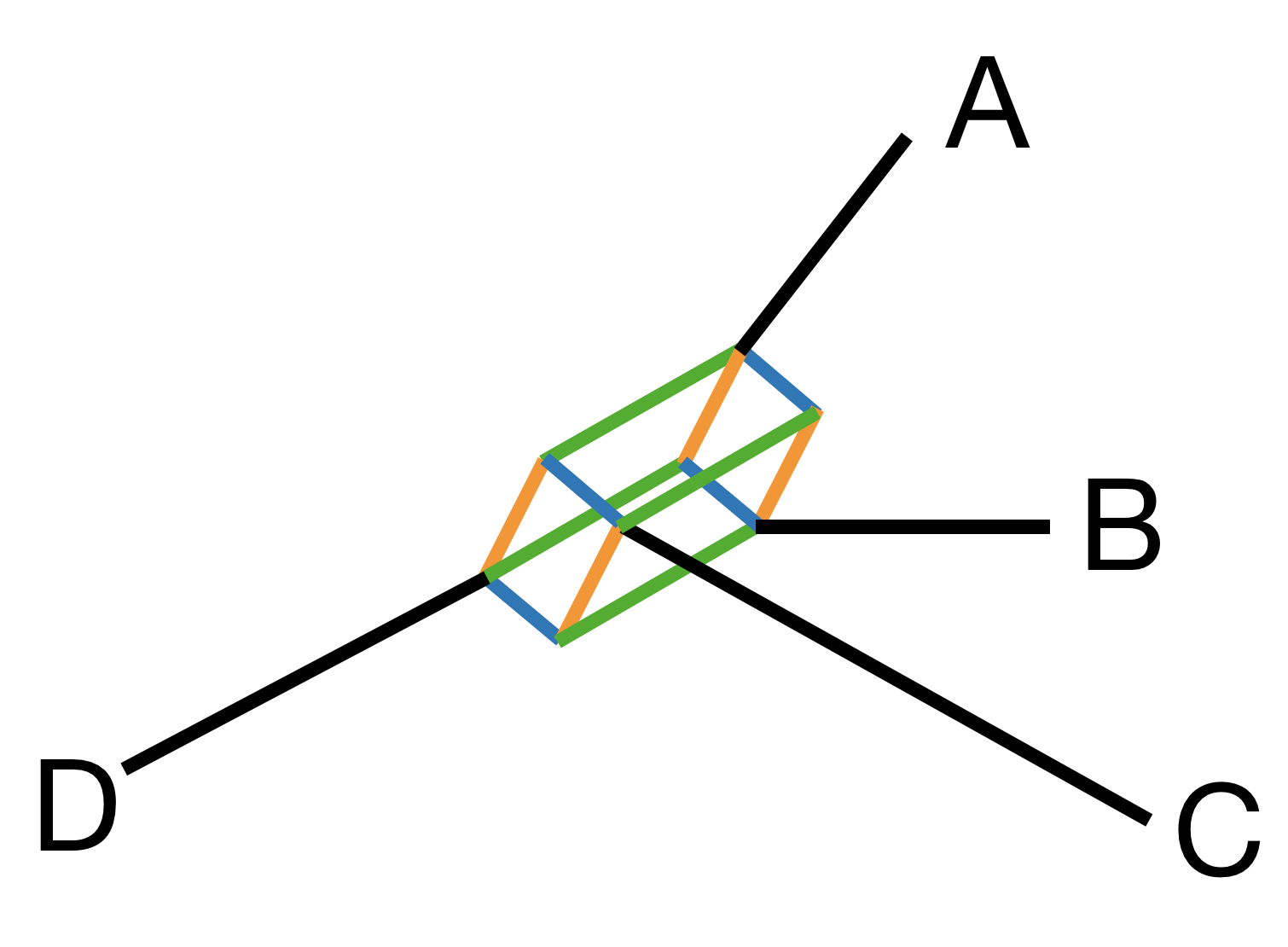}
    \includegraphics[width=0.4\linewidth]{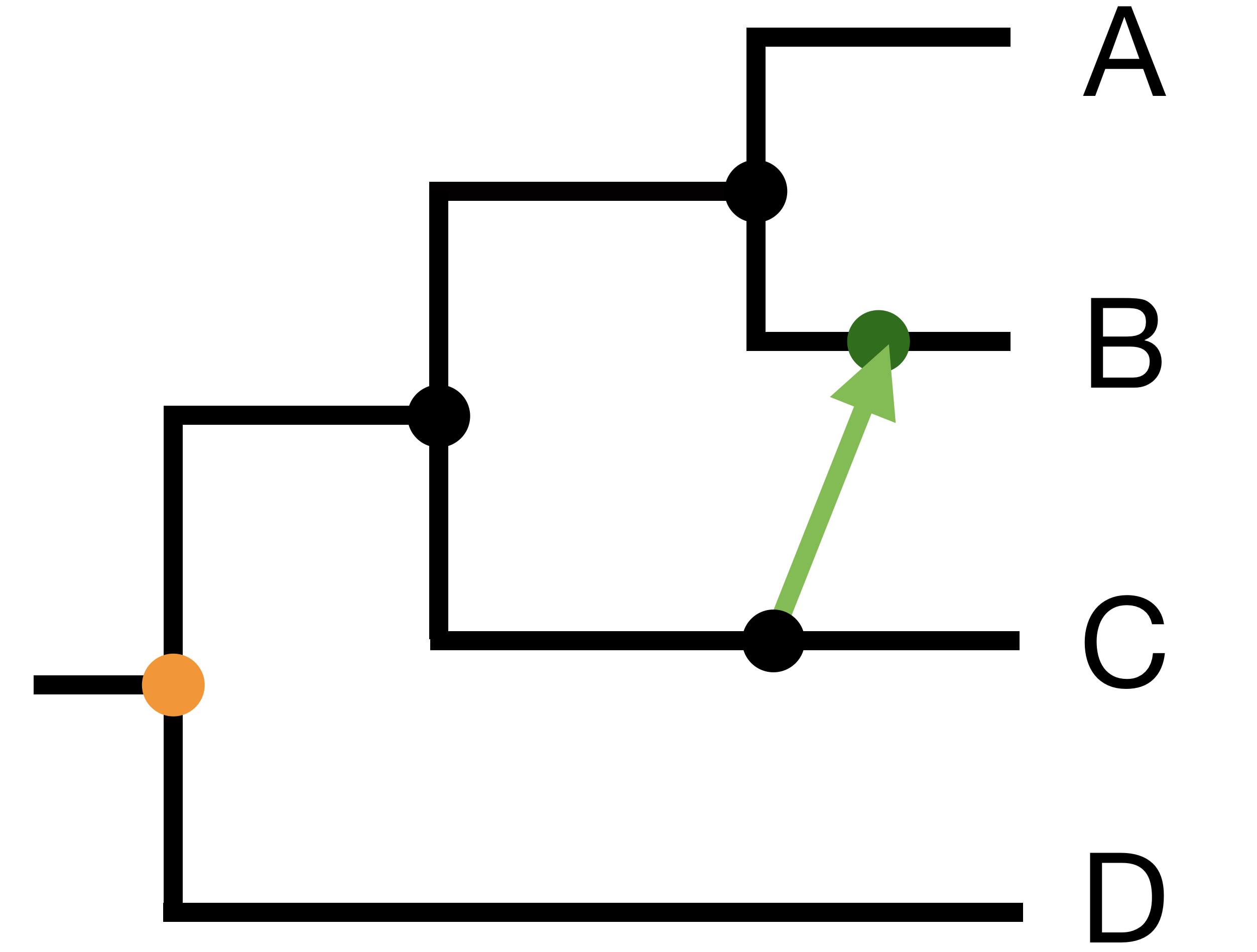}
    \caption{Left: Implicit (split) phylogenetic network where parallel edges (same color) correspond to a split. If we focus on one such split (green edges), we can collapse the alternative splits (blue and oranges edges) and get the $AB|CD$ quartet. Alternatively, if we focus on the blue split and we collapse the green and orange edges, we have the $AD|BC$ quartet. Each split network can be resolved into different trees by focusing on one way to collapse alternative splits. Right: Explicit phylogenetic network. Tree nodes (black) represent speciation events and hybrid nodes (green) represent reticulation events. The root node is shown in orange. A semi-directed network is obtained by removing the root node. Note that tree (black) nodes and the hybrid (green) node form what is denoted a \textit{hybridization cycle}.}
    \label{fig:split}
\end{figure}

At first glance, networks appear to be a straightforward extension of trees by the simple inclusion of a third type of node: \textit{hybrid nodes} with two parental incoming edges and one outgoing child edge (Figure \ref{fig:split} right). Hybrid nodes represent reticulate processes such as hybrid speciation, introgression or horizontal gene transfer. However, one immediate reality complicates our understanding of networks: there is in fact no universal definition of what internal nodes (or edges) represent in what is denoted (quite loosely) a \textit{phylogenetic network}. Indeed, scientists in different domains have used very different definitions of phylogenetic networks.

One such example is \textit{haplotype networks} \citep{templeton1992cladistic, excoffier1994using, fitch1997networks} and their \textit{median} \citep{bandelt1995mitochondrial, bandelt2000median} or \textit{median-joining} \citep{bandelt1999median} variants which represent the genealogical relationships among distinct haplotypes (unique combinations of linked alleles) observed within a population or species, typically inferred from non-recombining genetic loci. Nodes represent unique haplotypes, edges represent mutations between haplotypes, and internal nodes represent inferred ancestral haplotypes. These graphs can accommodate reticulations in the form of cycles, and thus, they extend tree models to account for recombination.
Despite not representing evolutionary relationships, haplotype networks continue to be used to represent mutational relationships, especially within recently diverged species or populations \citep{joly2007haplotype, leigh2015popart, paradis2018analysis} such as SARS-COV-2 \citep{forster2020phylogenetic, sanchez2020median, mavian2020sampling, chookajorn2020evolving}.


Another such example is implicit or \textit{split networks} \citep{bandelt1992split, bandelt1992canonical, huson2010phylogenetic}, also called abstract or data-display networks. 
A split is any bipartition of the taxa into two non-empty, disjoint and exhaustive taxon sets, and a split network represents every split as an array of parallel edges (Figure \ref{fig:split} left).
This non-tree-like representation is well-suited for visualizing reticulate evolution, 
but it cannot distinguish whether observed discordance is due to reticulation or to other processes such as ILS.
Indeed, internal nodes in split networks are devoid of biological meaning, and thus, they are better suited as data-display objects rather than as a representation of true evolutionary histories.  
Extensive software have been implemented to reconstruct split networks such as \texttt{SplitsTree} \citep{huson1998splitstree, huson2024splitstree}, \texttt{NeighborNet} \citep{bryant2004neighbor, bryant2023neighbornet}, \texttt{QNet} \citep{grunewald2007qnet} and \texttt{Quartet-Net} \citep{yang2013quartet}, and much theory 
has been developed in the reconstructability of split networks (i.e., conditions under which a network can be uniquely recovered) from different types of data \citep[e.g.]{gambette2012quartets, gambette2017uprooted, van2018leaf, francis2018tree, fischer2020space, fischer2021unrooted}.

In contrast to split networks, the nodes in \textit{explicit networks} represent a specific biological process: tree nodes (internal nodes with one parental edge and two children edges) represent speciation events and hybrid nodes (internal nodes with two parental edges and one child edge) represents reticulation events (Figure \ref{fig:split} right).
One could argue that the term \textit{phylogenetic networks} should be restricted to the case of \textit{explicit networks} as those are the ones that have an evolutionary interpretation. For the remaining of this work, we will refer to \textit{explicit phylogenetic networks} whenever we use the term \textit{phylogenetic networks}.

This perspective centers on explicit phylogenetic networks. Yet, even when restricting to explicit networks, the definition is not free of entanglements. Unlike species trees \citep{kubatko2026} where internal nodes have one clear biological interpretation (speciation), hybrid nodes in networks do not have a single biological meaning. Indeed, when inferring a phylogenetic network, hybrid nodes present a non-tree-like reticulate process, but no existing method is able to distinguish if the reticulate signal comes from hybridization or introgression or other gene flow process. In fact, depending on the method used to infer the phylogenetic network, we cannot even know if the signal is coming from other biological sources like duplications or losses, or even data errors or systematic noise.
Furthermore, unlike trees where there is not a ``wrong placement" of internal (speciation) nodes, hybrid nodes cannot be freely placed anywhere in the phylogeny due to risks of internal incongruences such as flow of genetic material contradicting temporal constraints imposed by outgroups or the root.
To further complicate matters, phylogenetic networks are classified into different groups by the number of hybrid nodes and their placement in the phylogeny (Figure \ref{fig:networks}). These groups differ widely in their ability to be estimated from  data and their biological interpretability, as we will see in the next sections.

\subsection{Exploring the untamed space of phylogenetic networks}

Just like a jungle on an alien planet, the space of phylogenetic networks is filled with unknown graphical creatures for which we know very little about their statistical detectability (can we estimate them accurately with the data we have?), their identifiability (could our data have been equally generated under two different networks?), or their biological interpretability (can this network represent an actual evolutionary history or it is just a mathematical object disconnected from reality?).

\begin{figure}
    \centering
    \includegraphics[width=\linewidth]{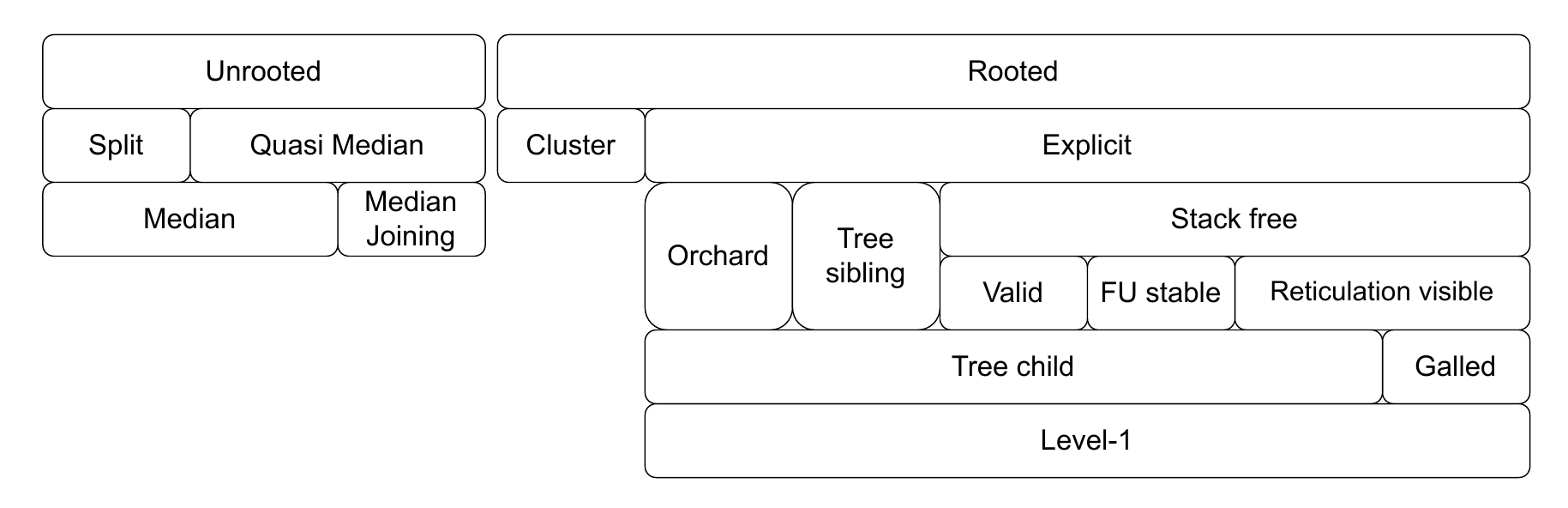}
    \includegraphics[width=\linewidth]{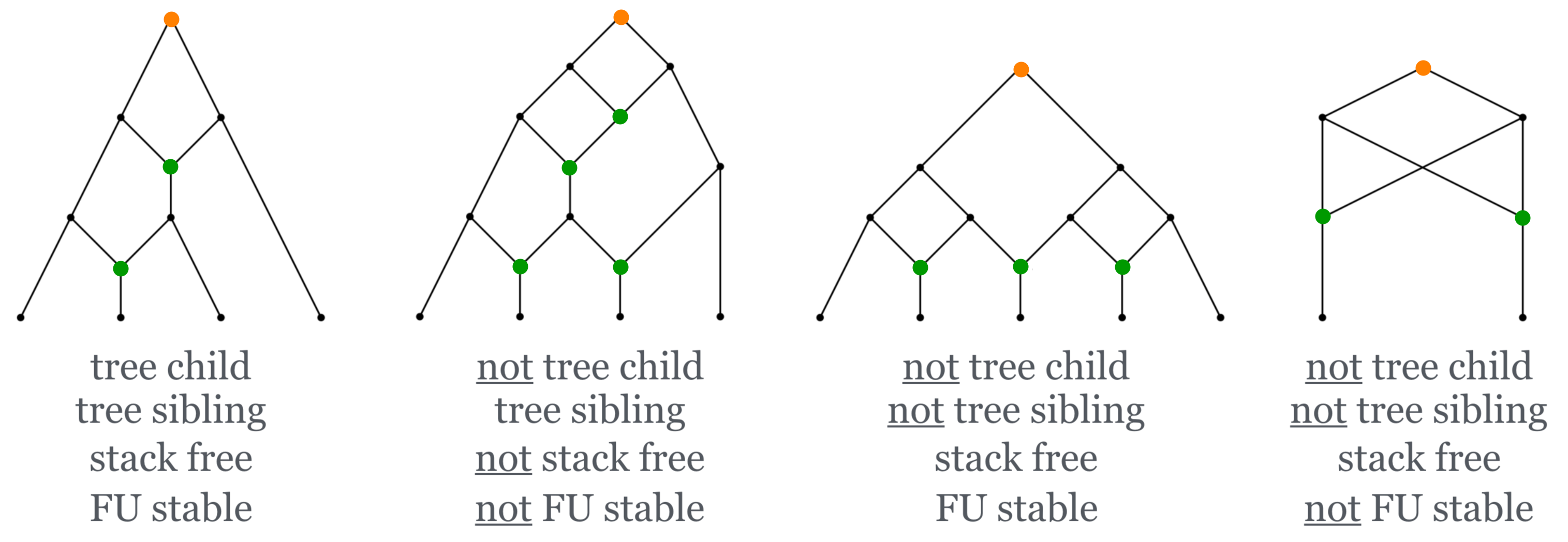}
    \caption{Top: Simplified schematic of network space showing only a subset of network classes. Belonging to a class works in an upward manner. For example, level-1 networks are necessarily tree child and galled; galled networks are reticulation visible, and all valid networks are stack free. Tree sibling networks are not necessarily stack free, though they could be, as in the case of tree-child networks that are both tree-sibling and stack-free. Bottom: Examples of networks in different classes, adapted from Fig.~4 in \cite{kong2022classes}. Tree nodes and hybrid nodes are shown in different colors. These examples are not intended to exhaustively define network classes, but to emphasize the diversity of network space and the fact that many important distinctions are not visually apparent hence the importance of mathematical definitions. For a more detailed schematic of network space, see Fig.~1 in \cite{huson2011survey} and Fig.~12 in \cite{kong2022classes}.}
    \label{fig:networks}
\end{figure}

Early attempts to make sense of this landscape took an almost taxonomic approach, defining rigorous classes of phylogenetic networks.
In a foundational textbook, \citet{huson2010phylogenetic} cataloged a variety of network types such as tree-based, tree-child, galled, and level-$k$ networks, each imposing different structural constraints (Figure \ref{fig:networks}). 
This figure intends to simply emphasize the richness and fragmentation of network space rather than to serve as a classification guide, and thus, to learn more about network classes, please refer to \citet{kong2022classes} who not only present a comprehensive list of network classes, but also show how some of these classes correspond to particular evolutionary mechanisms. This growing taxonomy, however, only partially tames the jungle: many network classes lack a clear biological interpretation, and for most, fundamental questions of statistical identifiability remain unanswered as we will see in future sections.


Most methods for inferring networks have specialized in \textit{level-1} networks.
Technically, the two incoming parental edges on a hybrid node create what is denoted as a \textit{hybridization cycle} (Figure \ref{fig:split} right), and in a level-1 network,
the hybridization cycles do not share edges. 
Level-1 networks have become the model organism of network space
as the class on which most theoretical results on identifiability and reconstructability have been established (see sections below).
Even methods that are capable of inferring more complex level-$k$ ($k>1$) networks are routinely evaluated only on level-1 networks.
So, ultimately, most of what we know about phylogenetic network inference is from the class of level-1 networks, and very few studies have tested how methods perform outside of this class \citep{cao2019practical}.

The network of life is unlikely to be level-1. As a field, we are used to making simplifications when representing biological processes with mathematical graphs. 
For example, the tree assumption of instantaneous isolation after speciation (hence represented by one lineage splitting into two at an internal node)
is a well-recognized biological idealization, for which alternative models exist but are not routinely implemented.
Restricting phylogenetic networks to level-1, however, is largely driven by current theoretical and computational constraints, rather than by a comparable biological rationale.
Thus, expanding our 
knowledge about identifiability and computational scalability to 
other classes of networks will allow us to include broader, more realistic evolutionary relationships of highly reticulate organisms in the network of life.

\section{Inferring the Network of Life}
\label{infer}

%

\subsection{Networks as trees with added hybrid arcs}

The evolution of network methods has followed an intuitive path. Given our field's success in inferring trees, one of the initial strategies to infer a network was to first infer a tree, and then test subsets of taxa for signal of reticulations.
These so-called ``hybrid detection methods" such as the D-statistics (ABBA-BABA) \citep{pattersonD} and similar D tests \citep{pease2015detection, leo2018d, hahn2019three, hibbins2019timing} became universally embraced to identify taxa of hybrid origin.
Unfortunately, recent work \citep{kong2021comparative, hibbins2022phylogenomic, frankel2023summary, bjorner2024detectability, koppetsch2024towards} has shown that 
D tests often exhibit elevated false discovery rates, erroneously classifying non-hybrid taxa as having hybrid origin, especially in cases of ancestral gene flow, ghost lineages \citep{tricou2022ghost, tricou2022ghostb}  or rate variability \citep{frankel2023summary}.

Not all hybrid detection methods are inspired by the D-statistics. 
\texttt{HyDe} \citep{blischak2018hyde} and \texttt{PhyloDiamond} \citep{wu2024ultrafast} use phylogenetic invariants under the multispecies coalescent (MSC) model, on site pattern frequencies and concordance factors respectively, to detect signal of reticulation on subsets of taxa, and
\texttt{MSCQuartets} \citep{rhodes2021mscquartets, allman2022gene} tests deviations from expectations under the MSC on a 4-taxon species tree. Likelihood ratio tests \citep{peng2024likelihood} have been introduced as more powerful alternatives to test for reticulation signal in site pattern frequencies.
Some of these tests, however, still show high false negative rates when involving extinct or unsampled ``ghost" lineages 
and under multiple hybridization events, implying that complex hybrid histories can attenuate rather than confuse the signal of gene flow \citep{bjorner2024detectability}.

Furthermore, all hybrid detection methods are affected by biases in tree inference under gene flow. Specifically, gene flow can generate anomalous gene trees that are overrepresented in the data, leading to inconsistent species tree estimation \citep{solis2016inconsistency, zhu2016light,long2018effect, dinh2025misspecification}. 
%
%
Gene flow can also lead to overestimated population sizes and underestimated species divergence times, and can result in lower posterior support for true clades, potentially yielding high confidence in an incorrect topology \citep{leache2014influence}.
Even low levels of gene flow 
can drastically distort species tree estimates, especially after rapid radiations
with consensus and distance-based methods outperforming MSC likelihood approaches that ignore gene flow
\citep{jiao2020impact}.

Machine learning has recently been introduced to distinguish
histories of speciations from histories of hybridization with high accuracy \citep{hibbins2022distinguishing}.
Once trained, machine‑learning models can be rapidly applied to new datasets
, yet they require labeled training datasets which, in phylogenetics, must be generated by simulation. Despite efforts to emulate realistic evolutionary scenarios across broad parameter ranges, such simulations inevitably introduce bias \citep{mo2024applications}.


\subsection{Networks as amalgamation of (sub)trees}

Early methods to reconstruct phylogenetic networks started as heuristic or distance-based with limited incorporation of explicit evolutionary models \citep{makarenkov2006phylogenetic}.
Tools like \texttt{T-REX} \citep{makarenkov2001t, legendre2002reconstruction, makarenkov2004phylogenetic, makarenkov2004modelling} were among the first to explicitly model reticulate evolutionary patterns using a distance-based approach.
%
However, because these algorithms are heuristic in nature, reticulation edges lack biological interpretation and may confound multiple sources of conflict, such as ILS or measurement error. 

Developments on the mathematical properties of splits, clusters and subgraph patterns within explicit networks gave rise to the formalization of theory and algorithms to build explicit networks from integral subparts.
For example, we can build explicit networks from 
\textit{triplets} \citep[3-taxon trees;][]{jansson2006algorithms, van2009constructing, huber2010practical, poormohammadi2014tripnet, reyhani2017rpnch},
\textit{quartets} \citep[4-taxon trees;][]{linz2020caterpillars, keijsper2014reconstructing},
\textit{binets} \citep[2-taxon subnetworks;][]{van2017binets, huber2017reconstructing},
\textit{trinets} \citep[3-taxon subnetworks;][]{huber2013encoding, van2014trinets, oldman2016trilonet, huber2017reconstructing, semple2021trinets, van2022algorithm},
\textit{quarnets} \citep[4-taxon subnetworks;][]{huber2018quarnet, huebler2019constructing, huber2024quarnets, frohn2025reconstructing}, 
\textit{clusters} \citep{van2010phylogenetic, wang2013bimlr, wang2013lnetwork},
and even \textit{ancestral profiles} \citep{erdHos2019class}. 

Substantial theoretical literature has characterized level-$k$ networks showing, for example, their uniqueness from triplets despite NP-hard reconstruction \citep{van2009uniqueness}, their decomposition into level-$k$ generators \citep{gambette2009structure}, and the possibility of polynomial-time reconstruction from clusters \citep{kelk2011elusiveness}.
Related work characterizes which level-1 networks can be encoded from trees, clusters, and triplets \citep{gambette2012encodings}, and shows how classes such as tree-child networks can be counted and enumerated for a fixed number of leaves and reticulations \citep{cardona2020counting}.
These results directly informed algorithms and software including \texttt{Dendroscope} \citep{huson2012dendroscope}, methods reconstructing level-1 networks from triplets \citep[\texttt{Lev1athan};][]{huber2010practical}, trinets \citep[\texttt{TriLoNet};][]{oldman2016trilonet}, and quarnets \citep[\texttt{Squirrel};][]{holtgrefe2025squirrel}, distance-based extensions of neighbor-joining \citep{willems2014new, yu2015distance, willems2018building}, and gene tree approaches such as \texttt{RF-Net} \citep{markin2022rf}.

All these algorithms are built upon solid mathematical and computational theory, yet they do not rely on explicit probabilistic models of data generation.
By amalgamating subgraphs into networks, these methods often fail to account for different sources of gene tree discordance or gene tree estimation error. Their heuristic setting makes it challenging to extend to different biological processes beyond reticulations, and thus, while these algorithms do build explicit networks that have an evolutionary interpretation, they are sometimes better suited as data-display objects just as split networks are.

\subsection{Networks as parameters in probabilistic models}


A shared dream of statistics and biology is to define a unified probabilistic model that accounts for \textit{all} possible biological processes involved in the generation of the observed data. 
Under the assumption of no recombination within loci, extending our tree probabilistic framework to networks only requires replacing the stochastic process that generates gene tree variability within a species tree with the corresponding process on a species network, given that the mutational process on gene trees that gives rise to sequences remains unchanged.

The multispecies coalescent (MSC) model on trees \citep{rannala2003bayes, rannala2020multispecies, kubatko2026} tracks the genealogical history of gene sequences going backwards in time and allows the estimation of species tree topology, divergence times, and population sizes. 
%
\cite{Meng2009} extended the MSC model to include hybridization events by incorporating a new numerical parameter: the inheritance probability $\gamma$ that models the probability of a single gene from the hybrid species to follow a specific path towards the most recent common ancestor. That is, in this model, if $B$ is the hybrid species with two sister species $A$ and $C$ (Figure \ref{fig:split} right), when a single gene from the hybrid species is selected for analysis, its most recent common ancestor occurs with taxon $C$ with probability $\gamma$ and with taxon $A$ with probability $1-\gamma$.
Upon selecting a species tree for a given gene, the coalescent process is applied to that tree, yielding probabilities for the various gene tree topologies.



This extended MSC model had one limitation: if there was more than one hybrid species (or multiple individuals sampled from the one hybrid species), all lineages below the hybrid node were forced to follow the same species tree. 
\cite{Yu2012a} extended this model to allow any hybrid lineage to follow any path, independently of other lineages, and this was further parametrized by \cite{fogg2023phylocoalsimulations} to include a $\rho$ parameter to determine the extent of correlation among lineages with $\rho=0$ corresponding to independent inheritance as in \cite{Yu2012a} and $\rho=1$ corresponding to common inheritance as in \cite{Meng2009}. 

This new probabilistic model that combined ILS and hybridization was 
denoted indistinguishably the \textit{multispecies network coalescent} (MSNC) or the \textit{network multispecies coalescent} (NMSC) and allows us to define a likelihood function to implement a maximum likelihood procedure that will traverse network space in search of the optimum network to fit the data. This methodology, originally implemented in \texttt{PhyloNet} \citep{Yu2014, wen2018inferring}, allows multilocus sequences or estimated gene trees as input data, later extended to Bayesian inference in \texttt{BEAST2} \citep{zhang2018bayesian} 
and in \texttt{PhyloNet} \citep{wen2016bayesian}. 

The shortage of sufficiently long genes to infer accurate gene trees in some taxonomic groups motivated the development of methods based on biallelic markers \citep{warnow2025advances}. Algorithms such as \texttt{PhyloNet} \citep{zhu2018bayesian, zhu2018inference}, \texttt{SnappNet} \citep{rabier2021inference}, and the pioneering \texttt{TreeMix} \citep{pickrell2012inference} infer networks from SNP data, although \texttt{TreeMix} is rarely considered within the realm of phylogenetic network methods given its focus on ``admixture graphs" (see Discussion).

Crucially, by defining a likelihood function (whose exact form depends on the data type and modeling assumptions), probabilistic approaches enable the inference of numerical parameters such as branch lengths and inheritance probabilities alongside network topology.
Focusing on numerical parameter inference,
\texttt{BPP} \citep{yang2015bpp} implements two models of gene flow under the MSC: 
1) the MSC with introgression (MSC-I or MSci) \citep{flouri2020bayesian, jiao2021multispecies}, which is the same as the MSNC, assumes episodic gene flow at discrete points in time, and 2) the MSC with migration \citep{jiao2021multispecies, flouri2023efficient} (MSC-M, also known as the isolation-with-migration or IM model \citep{hey2004multilocus}) which assumes continuous gene flow at a fixed rate after divergence.
While \texttt{BPP} does not currently infer the network topology,
it can test for gene flow and estimate numerical parameters such as migration or introgression rates, as well as species divergence times and population sizes.

The rise of likelihood-based approaches required not only the development of probabilistic models to define likelihoods (and priors), but also of computational tools to explore the vast space of networks efficiently.
Several studies have formalized network rearrangement moves that guarantee connectedness of network space. For example, \cite{huber2016spaces} generalized nearest-neighbor interchange (NNI) operations to level-1 networks, and \cite{gambette2017rearrangement} extended NNI and subtree pruning and regrafting (SPR) moves to include “horizontal” moves connecting networks of the same complexity and “vertical” moves connecting networks of different complexities. Other work has shown that connectedness can also be achieved using only tail moves \citep{janssen2018exploring} or head moves \citep{janssen2018heading}, and that tree-child networks are connected under a generalized SPR operation for a fixed number of hybridizations \citep{bordewich2017lost}.
Because the root is generally unidentifiable under likelihood-based models (see Statistical Identifiability section), many methods infer \textit{semi-directed networks}, in which hybrid edges and their descendants are directed while all other edges remain undirected. In this context, \cite{linz2023exploring} proposed a rearrangement move for semi-directed networks (cut edge transfer, CET) and showed that the space of semi-directed level-1 networks with a fixed number of reticulations is connected.

Although probabilistic models now routinely accommodate ILS and gene flow, extensions beyond these processes are rare.
The dream of integrating several (\textit{all}) biological processes into one probabilistic framework is complicated for statistical and computational reasons, as we will see in the next section.
Network models described so far account for ILS and gene flow; none can tackle allopolyploidization.
Polyploids can arise through hybridization between distinct species, resulting in distinct subgenomes that are allopolyploids, but this biological process is ignored by most existing network models that assume species are diploids.
One notable exception is \texttt{AllopNet} \citep{jones2013statistical, jones2017} which infers allopolyploid phylogenetic networks in the presence of ILS, but is restricted to a single hybridization event between 2 diploid species resulting in a genomic allotetraploid. 
Further efforts are required to continue to develop probabilistic models that encompass a broader array of biological processes, while rigorously addressing the inevitable statistical and computational challenges, some of which we describe next.

\section{The challenges of network inference}
\label{challenges}

Built upon biologically-informed probability models, likelihood-based network inference methods have established as accurate methodologies to infer reticulate evolutionary histories. These methods, however, 
have been confronted with two primary challenges.
First, the prohibitive computational cost of full likelihood evaluation on phylogenetic networks rendered these methods impractical for large empirical datasets (lack of scalability).
Second, as the complexity of the probability models increases, it becomes more likely that different parameter settings produce similar patterns in the data such as gene trees or site frequencies (lack of identifiability). 
For example, for the case of network topologies, as we increase the network complexity (number and placement of hybridizations), distinct network topologies can generate identical data (e.g. gene tree or site pattern distributions), rendering them statistically indistinguishable.

\subsection{Computational scalability}

\paragraph{Dividing and conquering.}
Likelihood computation can be accelerated by decomposition into smaller, more tractable components.
\texttt{SNaQ} \citep{Solis-Lemus2016-ux} pioneered a composite (pseudo) likelihood framework by decomposing the full likelihood into quarnets, a strategy also implemented in \texttt{PhyloNet} \citep{yu2015maximum} and in \texttt{PhyNEST} \citep{kong2025inference}. 
However, composite likelihood strategies alone are not likely to achieve the scalability of tree methods that can routinely handle thousands of taxa \citep[e.g.]{one2019one}. Even when likelihood optimization is indeed accelerated, the optimization must still traverse network space which grows super-exponentially with the number of taxa. 
%

\texttt{FastNet} \citep[polynomial complexity;][]{hejase2018fastnet} and later \texttt{InPhyNet} \citep[linear complexity;][]{kolbow2025method} provide a decomposition strategy to split the full problem (network inference on the whole set of taxa) into smaller subproblems (network inferences on a subset of taxa), and then merge solutions of the subproblems into a solution for the full problem. \texttt{InPhyNet} is the first network reconstruction method to achieve the same levels of number of taxa as trees by building the phylogenetic network for the 10,000 plant genomes \citep{one2019one}.

Other divide-and-conquer strategies combine probability models and heuristic approaches. For example, \texttt{NANUQ} \citep{allman2019nanuq} infers level‑1 networks by analyzing quartets displayed on gene trees. Its updated version \texttt{NANUQ+} \citep{allman2024nanuq} first infers the \textit{tree of blobs} \citep{allman2023tree}: a tree where hybridization biconnected components (blobs) are collapsed into multifurcating nodes, and then resolves multifurcations as hybridization events. \texttt{TINNiK} \citep{allman2024tinnik}, used within \texttt{NANUQ+}, infers trees of blobs from gene quartet distributions under the MSNC. Similarly, \texttt{Squirrel} \citep{holtgrefe2025squirrel} infers candidate trees of blobs and then resolves them into level-1 networks.
Some other methods simplify computation by ignoring biological processes, such as \texttt{NetRax} \citep{lutteropp2022netrax} that performs maximum likelihood network inference without modeling ILS.

\paragraph{Bypassing optimization.}
Algebraic geometry provides a theoretical framework that enables faster and more efficient phylogenetic network inference through the use of algebraic invariants: polynomial equations that model parameters have to satisfy for internal consistency. For example, \textit{concordance factors} \citep{baum2007concordance} represent the proportion of genes that support each of the three possible quartet representations for a 4-taxon subset: $ab|cd, ac|bd, ad|bc$, and these three concordance factors need to sum up to one. Thus, the equation $CF_{ab|cd} + CF_{ac|bd} + CF_{ad|bc} = 1$ needs to be satisfied, and thus, the polynomial equation $f(x,y,z) = x+y+z-1$ for $x=CF_{ab|cd}, y=CF_{ac|bd}, z=CF_{ad|bc}$ is an algebraic (phylogenetic) invariant.

Phylogenetic invariants have been used to infer 4-taxon networks under a variety of models of evolution
such as Cavender–Farris–Neyman \citep{cummings2024invariants} and Jukes–Cantor \citep{Barton2022} for level-1 networks, with subsequent extensions to level-2 networks under Jukes–Cantor \citep{ardiyansyah2021distinguishing, englander2025identifiability}.
\cite{martin2023algebraic} extends to any 4-taxon semi-directed network under the Jukes-Cantor and the Kimura 2-parameter models, and \cite{frohn2024invariants} under the random walk 4-state Markov model.
These invariants account for the mutation process of sequences under networks, but not for gene tree variability under the coalescent model. 
In contrast, \citet{wu2024ultrafast} use phylogenetic invariants defined on concordance factors under the MSNC, which enables inference of level-1 networks for $4<n<9$ taxa and serves as a hybrid detection tool for larger $n$.

Because these methods evaluate polynomial equations instead of exhaustively traversing network space, they achieve greater computational efficiency than traditional likelihood-based approaches. However, defining invariants becomes increasingly challenging as network size grows, limiting their applicability primarily to smaller networks where likelihood-based methods remain tractable. Future research could focus on integrating these strategies, for example, by using invariant-based approaches to preselect candidate networks for subsequent likelihood optimization and parameter estimation.

\subsection{Statistical identifiability}

Intuitively, two networks that generate the same data patterns would not be distinguishable from that data alone.
Therefore, it is essential to determine the model assumptions for which a specific data type uniquely identify a single network topology and its numerical parameters.
That is, we need to assess \textit{statistical identifiability} of models on networks.
These results depend on both the form of the data (gene trees, quartet concordance factors, or sequence data), and the underlying modeling assumptions such as substitution models and whether gene tree variation is explicitly modeled under the MSNC.
Statistical identifiability transcends purely theoretical considerations since understanding which local network features are most readily identifiable (and with which type of data) can help guide the design of sampling strategies that maximize detectability of reticulations.

\paragraph{Building intuition: early studies on identifiability.}
\cite{Solis-Lemus2016-ux} began the investigations on identifiability (or more precisely, detectability) of hybridization cycles on level-1 networks under the MSNC from quartet concordance factors (CFs). 
They found that the root is not identifiable which motivated further development of theory and methods on \textit{semi-directed} networks. 
This early study focused on detectability (that is, whether hybridization events could be statistically detected from quartet CFs) rather than on uniquely identifying the underlying network, and showed that our ability to detect hybridization cycles (from quartet concordance factors under MSNC) depends on the number of nodes in that cycle, with 2-cycles (cycles with 2 nodes) never detectable; 5-cycles (or bigger) always detectable, and 3- and 4-cycles detectable under certain taxon sampling conditions.

At the same time, \cite{pardi2015reconstructible} were examining network identifiability under a no-ILS assumption, which allows the full network likelihood to be decomposed into the likelihoods of its displayed trees (each obtained by removing one hybrid edge per hybridization). They showed that canonical networks, those 
in which no node has multiple incoming edges and a single outgoing edge,
are generally identifiable, whereas certain configurations with stacked hybridizations (hybrids giving rise to hybrids) remain indistinguishable.
\cite{zhu2017displayed} later showed that networks deemed not identifiable under no ILS by \cite{pardi2015reconstructible} are actually distinguishable under the MSNC (which accounts for ILS) when more than two lineages are sampled below the hybrid.
A central intuition emerging from these initial investigations is that taxon sampling incorporating multiple alleles per species, and gene tree variability due to ILS can help network distinguishability. Moreover, even when full networks remained indistinguishable, certain features such as their canonical, semi-directed, 2-cycle free or triangle (3-cycle) free forms may still be identifiable.

%


\paragraph{Level-1 networks are identifiable across models and data types.}
\cite{banos2019identifying} proved full (generic) identifiability of level-1 semi-directed networks under the MSNC from gene quartet distributions with cycles of size at least 4, later extended to 3-cycles by \cite{allman2024identifiability}. Generic identifiability means for all values of numerical parameters except some in a set of measure zero (\textit{almost all} values of numerical parameters).
These works allow us to conclude that level-1 semi-directed networks (topology and numerical parameters) are generically identifiable under the MSNC from gene quartet distributions except in the presence of 2‑cycles and, in certain cases, hybrid nodes and adjacent edges within 3‑cycles. 

Similar results have been proven under other models and data types.
Level-1 networks with cycles of size at least 4 were proven identifiable 
under the Jukes–Cantor substitution model for sequence data \citep{Gross2018}, later extended to the Kimura 2- and 3-parameter models \citep{gross2021distinguishing}, with subsequent work demonstrating that even the number of reticulations in a level-1 network is identifiable under Jukes–Cantor \citep{englander2025identifiability}.
Level-1 ultrametric networks, under MSNC combined with a mixture of General Time-Reversible evolution models for sequence data, are identifiable from logdet distances when cycles have size at least 4 \citep{allman2022identifiability}.
\cite{xu2023identifiability} showed that level‑1 networks with positive internal edge lengths are identifiable from average pairwise distances, except for local ambiguities in small (4-) cycles where hybrid edge orientation and adjacent branch lengths and inheritance parameters are not identifiable.

\paragraph{Some networks are identifiable beyond level-1.}
Stepping outside the class of level-1 networks requires strong assumptions on the network structure to guarantee identifiability. Specifically, it seems that ``hybridizations on top of hybridizations" can be quite problematic. Scenarios where hybrid nodes have hybrid nodes as children or there is a path from the root to the tips that traverses multiple hybridizations cannot be easily detected from different data types.
Once we restrict these scenarios, we can have identifiability results outside the level-1 class.
%
Namely, \cite{allman2025beyond} proved identifiability of more general networks from quartet concordance factors under MSNC. Ruling out small (2- or 3-) cycles, a network is identifiable if it is \textit{galled} and all hybridization blobs are \textit{tree-child}, meaning that hybridization events are isolated rather than layered on top of one another. Intuitively, galled networks restrict each hybridization to a single undirected (tree) cycle formed by two disjoint paths, while the tree-child condition ensures that every internal node has at least one descendant that follows a tree-like lineage, thereby ruling out stacked hybridizations in which a hybrid node gives rise to another hybrid node.
Their study also corroborated what we knew from early identifiability studies: multiple samples per taxon aid in network identifiability compared to single samples.

\cite{englander2025identifiability} showed that level-2, triangle (3-cycle) free and strongly tree-child networks are identifiable from sequence data under the Jukes–Cantor model, assumptions that again exclude stacked hybridizations.
%
The recurring exclusion of stacked hybridizations in identifiability results may indicate that the
canonical network classes (which does not allow stacked hybridizations) may capture the identifiable structure even in complex reticulation scenarios.

\paragraph{General network features emerge as identifiable across models and data types.}
Beyond topology and numerical parameters, several network features are also identifiable under different models and data types.
%
For example, \cite{allman2023tree} proved that the unrooted tree of blobs can be determined from the distribution of gene quartet topologies under the MSNC
and is also identifiable under Jukes-Cantor and the Kimura-2-Parameter substitution models from sequence data \citep{englander2025identifiability}.
A refinement of this tree, the distance split tree, is identifiable from pairwise distances \citep{xu2023identifiability}.
As mentioned before, the identifiability of the tree of blobs has opened the door to multiple inference network methodologies that first infer a tree of blobs to later identify the hybridization blobs \citep{allman2024nanuq, holtgrefe2025squirrel}.
Other features, such as the circular order of subnetwork attachments to a blob, are also identifiable 
\citep{rhodes2025identifying, englander2025identifiability}.

\paragraph{Proven cases of unidentifiability.}
Because the root is not identifiable across models and data types \citep{Solis-Lemus2016-ux, banos2019identifying, xu2023identifiability}, many methods these days infer semi‑directed networks. Yet, as evolutionary histories inherently require a root for interpretation, these networks must be rooted post hoc, a process that can introduce interpretation challenges (see Discussion).
Small cycles of length 2 or 3 tend to also be unidentifiable across models and data types \citep{Solis-Lemus2016-ux, gross2021distinguishing, xu2023identifiability, englander2025identifiability}.
Some 3-cycles in particular can create \textit{anomalous} networks \citep{ane2024anomalous} where an unrooted tree displayed by the network may have lower probability under the MSNC than a tree not displayed. Such anomalies complicate inference from gene tree data.
Biologically, simply discarding small cycles from networks is not ideal. Highly prevalent gene flow events in the network of life are expected to appear in small cycles. For example, introgression or migration between sister extant taxa is by definition a 3-cycle. These reticulation events are common among reticulate organisms, yet most of our current theory rules them out as unidentifiable.
Numerical parameters around hybridizations can also be hard to infer, especially around small cycles \citep{xu2023identifiability, allman2024identifiability}.

In terms of network classes, it has been proven that level-2 networks are not identifiable from average distances, not even generically \citep{xu2023identifiability}.
Looking ahead and borrowing ideas outside of probabilistic modeling, \cite{huber2024quarnets} recently showed that level-3 networks are not encoded by quarnets, and \cite{englander2025identifiability} demonstrated that any two binary semi-directed phylogenetic networks are only distinguishable if they do not display exactly the same quartets. Together, these results suggest that level-3 networks could be challenging to identify under some settings.
Stacked hybridizations are also hard or impossible to detect \citep{englander2025identifiability}. This might motivate the focus on canonical (or somehow reduced) networks \citep{pardi2015reconstructible, francis2021normalising, heiss2025transformations} which would prompt us to look beyond binary networks.

As we move farther from level-1-ness, more restrictions will need to be imposed on the networks to prove statistical identifiability, and this investigation will shape the type of network we can infer from different data types.
Evolutionary paths from root to extant species that involve multiple hybridization events are hard to infer with our current models and data types. This suggests exploring new types of data, different sampling strategies, or simplified network structures that focus on the most identifiable features. Just as the field quickly moved into semi-directed networks as the estimable object of interest, as we move outside level-1-ness, we could also think of canonical or other forms of reduced networks as our identifiable parameter of interest.


\section{Discussion}

This perspective centers on explicit phylogenetic networks inferred 
under an assumed probabilistic model for data generation.
This choice was subjective. 
As a statistician, I favor probabilistic approaches. As an evolutionary biologist, I favor biological interpretability. However, 
the field of phylogenetic networks is broad and rich, and any of the subtopics I tangentially discussed could be the focus of its own perspective piece. 

This perspective has also focused on the methods to build phylogenetic networks rather than on the extensive body of research on reticulations across the diverse array of organisms comprising the network of life (a small subset of those studies mentioned in the Introduction).
This perspective is also not meant to guide a novice into the field of phylogenetic networks. Other reviews on phylogenetic networks (better suited as introductions to the topic) have been published over the years \citep{linder2004network, morrison2005networks, huson2006application, makarenkov2006phylogenetic, nakhleh2010evolutionary, bapteste2013networks, degnan2018modeling, Elworth2019, blair2020phylogenetic, kong2025inference}.

I conclude this Discussion by highlighting additional challenges, each also deserving a review of their own.


\paragraph{Interpretations.}
Interpreting trees, even unrooted trees, is relatively straightforward. A tree represents a sequence of speciation events and is rarely subject to ambiguous visualization. It is not the same for networks. The manner in which networks are plotted (many times a subjective decision by the plotting software) can impact the way in which they are interpreted. Among the plotting software for phylogenetic networks, we can highlight \texttt{Dendroscope} \citep{huson2012dendroscope}, \texttt{PhyloPlots} \citep{phyloplots}, \texttt{tanggle} \citep{tanggle}, and \texttt{ggret} \citep{specht2025ggret}. Each software has a default way to plot phylogenetic networks.

\begin{figure}
    \centering
    \includegraphics[width=\linewidth]{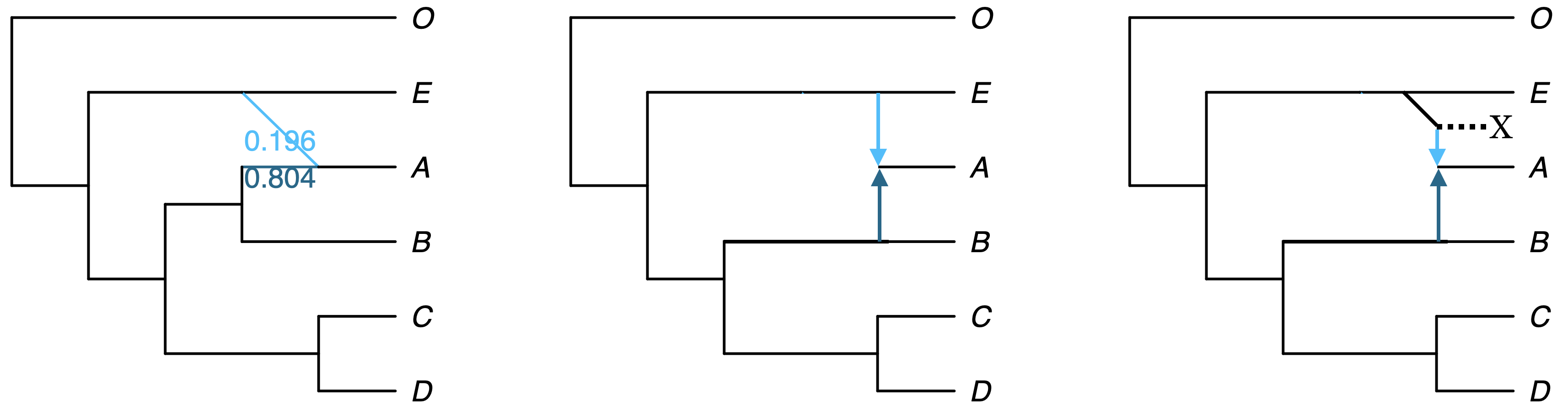}
    \caption{Same network topology plotted in three different ways to illustrate challenges in network interpretability. Left: This network seems to represent gene flow from the ancestral population of $E$ into the ancestral population of $A$ (figure plotted with \texttt{PhyloPlots} \citep{phyloplots}). Center: This network seems to represent hybrid speciation that gives rise to $A$. Right: This network represents gene flow from unsampled or extinct taxon $X$ into $A$. Mathematically, these three networks are identical.}
    \label{fig:root}
\end{figure}

Figure \ref{fig:root} shows three ways to plot the same network topology, but with different biological interpretations. On the left, the network seems to represent introgression or horizontal gene transfer, that is, a reticulate process in which there is gene flow from one population (ancestor of $E$) to another population (ancestor of $A$). The center figure seems to indicate that taxon $A$   arises through a process of hybrid speciation. Unlike the left network, this center network does not have estimates for the inheritance probabilities to aid in the interpretation, and the manner in which it is plotted could suggest equal contribution on both hybrid edges. The figure on the right represents gene flow from an unsampled or extinct taxon $X$ into taxon $A$. It should be emphasized that, from a mathematical standpoint, the three network topologies are identical. Any perceived differences arise solely from artifacts of their graphical representation. 


So, how can we interpret inferred networks? First, numerical parameters are important components of the inferred network. In trees, the branch lengths do not alter the pattern of speciation events, so a tree topology (even without branch lengths) can still be biologically interpreted. For networks, branch lengths and inheritance probabilities are necessary to provide biological meaning. Branch lengths in calendar time could provide information about the time points of the origin and the target of gene flow, and whether these two time points coincide. Estimated branch lengths of $0$ for the hybrid edges are  represented by perpendicular lines as in Figure \ref{fig:root} center. Given that gene flow should happen between contemporary populations, a hybrid edge with positive length could provide evidence of hybridization with an unsampled or extinct taxon (as in Figure \ref{fig:root} right). The inheritance probability can also provide information on whether the gene flow event is introgression (small $\gamma$) or hybrid speciation ($\gamma$ close to $0.5$). We should emphasize, however, that the
interpretation of inheritance probabilities depends on whether the model represents episodic or continuous gene flow. 
In a continuous gene flow model, it is possible to 
have values of $\gamma$ near 0.5 that do not invoke hybrid speciation \citep{kong2025pnas}, so caution is needed to interpret $\gamma$ as the sole source of evidence for hybrid speciation versus introgressive hybridization.

When interpreting networks, it is also important to understand the constraints under which the inference method operates. For example, a method that can only infer level-1 networks will have a limit on the number of hybridization events it can place on a phylogeny. Selecting $h=5$ hybridization events to be estimated, but obtaining a network with only $h=3$ does not mean that there is biological or even statistical evidence for only 3 hybridization events instead of 5. The method might have not been able to place more hybridization events given the level-1 constraint. Some methods can place constraints on hybrid edges as well, sometimes due to identifiability issues, forcing the lengths to be equal to $0.0$, and thus, they cannot be used as evidence for contemporary gene flow or hybrid speciation as they are an artifact of the optimization constraints.


\paragraph{Robustness.}
Identifiability results clarify what is theoretically detectable from data, but they do not address how network inference methods behave when model assumptions are violated.
Despite extensive work on the robustness of tree methods under gene flow \citep[e.g.]{solis2016inconsistency, zhu2016light,long2018effect, dinh2025misspecification, scornavacca2025inferring}, analogous studies for phylogenetic networks remain scarce. 
Existing work shows that data errors and model misspecification can spuriously increase inferred network complexity: sequencing and gene-tree errors may introduce non-biological reticulations \citep{tamiji2025impact}, while ignoring gene-tree uncertainty or assuming homogeneous substitution rates can have similar effects \citep{cao2024impact}. Practical evaluations of software such as \texttt{PhyloNet} have begun to translate these findings into guidance for users \citep{cao2019practical}. 
More studies are needed to assess the robustness of our network inference methodologies as these methods tend to have several topological and numerical assumptions that could be violated in empirical data.

\paragraph{Distances and consensus.}
Computational operations that are straightforward on trees such as calculating distances or dissimilarities, and finding consensus graphs become substantially more complex when extended to networks.
Key challenges arise because distance functions that work well within one class of networks often fail to behave as proper distances in other classes, and commonly used units to build distances or consensus such as splits or clades do not map to unique edges in networks. 
More work is needed to find better distance functions that properly capture the notion of similarity among phylogenetic networks \citep[e.g.]{cardona2008metrics, cardona2008distance, cardona2008metrics2, nakhleh2009metric, huson2010phylogenetic, moulton2017cubic, yakici2022phylogenetic, maxfield2025dissimilarity}, as well as algorithms for computing consensus networks from collections of networks (not just trees) \citep{huber2023computing} to support Bayesian inference and other uncertainty quantification frameworks.

\paragraph{Traits.}
Phylogenetic comparative methods such as phylogenetic regression, phylogenetic ANOVA, ancestral trait reconstruction, and Pagel's $\lambda$ test of phylogenetic signal study the distribution of quantitative traits among related species. 
\citet{jhwueng2015trait, bastide2018phylogenetic, karimi2020reticulate, teo2023accounting, teo2025leveraging} present comparative methods where the phylogenetic correlation structure of a trait is governed by a network.
Existing models, implemented in \texttt{PhyloTraits} \citep{phylotraits}, already allow for network-specific tests such as transgressive evolution. However, substantial work remains to match the breadth of trait models available for trees.

\paragraph{A network by any other name.}
As noted earlier, the term 'phylogenetic network' can have different meanings on different domains.
It turns out that the converse is also true: there exist terminologies that refer to objects that mathematically are phylogenetic networks under different names.
\textit{Admixture graphs} \citep{pickrell2012inference, lipson2013efficient, loh2013inferring, peter2016admixture, lipson2020applying, kamm2020efficiently, nielsen2023bayesian, maier2023limits}, highly popular in human population genetics, are graphical models that represent divergence and admixture events in a population's history.
Although mathematically equivalent, admixture graphs and phylogenetic networks have evolved in parallel fields, with largely non-overlapping theoretical foundations and methodologies.

A similar isolation exists between phylogenetic networks and \textit{ancestral recombination}
\citep{bloomquist2010unifying, vaughan2017inferring, mostowy2017efficient, deng2021distribution, yc2022evaluation, muller2022bayesian}
or \textit{reassortment} 
\citep{yurovsky2010flurf, muller2020bayesian, barrat2022treeknit, stolz2022joint}
\textit{graphs}, both ambiguously referred to as ARGs.
While these ARGs capture gene-level reticulate processes such as recombination or reassortment, their graphical structure is mathematically equivalent to that of a phylogenetic network.

While the theoretical and methodological paths of these network objects have diverged, perhaps it is time to recombine them by focusing on their shared characteristics, and to exploit the substantial opportunities for cross-fertilization that remain unexplored.
Network theory aims to understand the mathematical objects that we denote networks, to study their statistical, mathematical and computational properties, and to develop efficient, accurate and scalable ways to infer them from available data. 
Although different domains rely on different empirical data and probabilistic models tied to specific biological processes, the inherent graphs are networks that share commonalities that warrant a unified and comprehensive investigation.



\paragraph{}
Just like species on Earth, our methodologies evolve. 
The more deeply we study reality, the more anomalies we find, and the bubble bursts: our current methodologies cannot handle some realities. 
A tree structure is not enough to represent
the evolutionary history of all species on Earth.
%
As we improve our network models and better ways to extract information about reticulate processes from empirical data, we can 
help transition to a network thinking era, a new paradigm that is not meant to overrule the idea of a tree of life, but rather to expand it into a broader picture of evolution.

\section*{Funding}
This work was partially supported by the National Science Foundation (NSF DEB-2144367 to CSL).

\section*{Acknowledgements}
I want to thank Chris Simon and Stacey Smith for inviting me to contribute to the 75th anniversary edition of the journal of Systematic Biology as well as Bob Thomson, Paul Bastide and Laura Kubatko for the thorough comments that greatly improved this work.
It was an honor to acknowledge the contributions of past and present researchers who dared venture into the vast network space, laying the methodological foundations that make it possible to reveal the hidden patterns of reticulate evolution in data.
I am also very grateful for the continuous support throughout my career of the Society of Systematic Biologists.

\section*{Disclosure statement}
No conflicts of interest to disclose.

\section*{Data Availability Statement}
No data associated with this perspective article.

\bibliographystyle{unsrtnat}
\bibliography{references}

@book{Kuhn1962,
  author = {Kuhn, Thomas S.},
  title = {The Structure of Scientific Revolutions},
  year = {1962},
  publisher = {University of Chicago Press},
  address = {Chicago, IL},
}

@article{Meng2009,
author = {Meng, Chen and Kubatko, Laura Salter},
doi = {10.1016/j.tpb.2008.10.004},
isbn = {1096-0325 (Electronic)$\backslash$r0040-5809 (Linking)},
issn = {1096-0325},
journal = {Theoretical population biology},
month = {mar},
number = {1},
pages = {35--45},
pmid = {19038278},
title = {{Detecting hybrid speciation in the presence of incomplete lineage sorting using gene tree incongruence: a model.}},
volume = {75},
year = {2009}
}

@article{Yu2014,
author = {Yu, Yun and Dong, Jianrong and Liu, Kevin J and Nakhleh, Luay},
doi = {10.1073/pnas.1407950111},
journal = {PNAS},
number = {46},
pages = {16448--16453},
title = {{Maximum Likelihood Inference of Reticulate Evolutionary Histories}},
volume = {111},
year = {2014}
}

@article{Yu2012a,
author = {Yu, Yun and Degnan, James H. and Nakhleh, Luay},
doi = {10.1371/journal.pgen.1002660},
isbn = {1553-7390},
issn = {1553-7404},
journal = {PLoS genetics},
month = {jan},
number = {4},
pages = {e1002660},
pmid = {22536161},
title = {{The probability of a gene tree topology within a phylogenetic network with applications to hybridization detection.}},
volume = {8},
year = {2012}
}

@article{kong2022classes,
  title={Classes of explicit phylogenetic networks and their biological and mathematical significance},
  author={Kong, Sungsik and Pons, Joan Carles and Kubatko, Laura and Wicke, Kristina},
  journal={Journal of Mathematical Biology},
  volume={84},
  number={6},
  pages={47},
  year={2022},
  publisher={Springer}
}

@article{tricou2022ghost,
  title={Ghost lineages highly influence the interpretation of introgression tests},
  author={Tricou, Th{\'e}o and Tannier, Eric and de Vienne, Damien M},
  journal={Systematic Biology},
  volume={71},
  number={5},
  pages={1147--1158},
  year={2022},
  publisher={Oxford University Press}
}

@article{tricou2022ghostb,
  title={Ghost lineages can invalidate or even reverse findings regarding gene flow},
  author={Tricou, Th{\'e}o and Tannier, Eric and de Vienne, Damien M},
  journal={PLoS biology},
  volume={20},
  number={9},
  pages={e3001776},
  year={2022},
  publisher={Public Library of Science San Francisco, CA USA}
}

@article{hibbins2022distinguishing,
  title={Distinguishing between histories of speciation and introgression using genomic data},
  author={Hibbins, Mark S and Hahn, Matthew W},
  journal={BioRxiv},
  pages={2022--09},
  year={2022},
  publisher={Cold Spring Harbor Laboratory}
}

@article{leo2018d,
  title={D GEN: a test statistic for detection of general introgression scenarios},
  author={Leo Elworth, Ryan A and Allen, Chabrielle and Benedict, Travis and Dulworth, Peter and Nakhleh, Luay},
  journal={BioRxiv},
  pages={348649},
  year={2018},
  publisher={Cold Spring Harbor Laboratory}
}

@article{blischak2018hyde,
  title={HyDe: a Python package for genome-scale hybridization detection},
  author={Blischak, Paul D and Chifman, Julia and Wolfe, Andrea D and Kubatko, Laura S},
  journal={Systematic Biology},
  volume={67},
  number={5},
  pages={821--829},
  year={2018},
  publisher={Oxford University Press}
}

@article{pease2015detection,
  title={Detection and polarization of introgression in a five-taxon phylogeny},
  author={Pease, James B and Hahn, Matthew W},
  journal={Systematic biology},
  volume={64},
  number={4},
  pages={651--662},
  year={2015},
  publisher={Oxford University Press}
}

@article{hahn2019three,
  title={A three-sample test for introgression},
  author={Hahn, Matthew W and Hibbins, Mark S},
  journal={Molecular biology and evolution},
  volume={36},
  number={12},
  pages={2878--2882},
  year={2019},
  publisher={Oxford University Press}
}

@article{kong2021comparative,
  title={Comparative performance of popular methods for hybrid detection using genomic data},
  author={Kong, Sungsik and Kubatko, Laura S},
  journal={Systematic Biology},
  volume={70},
  number={5},
  pages={891--907},
  year={2021},
  publisher={Oxford University Press}
}

@article{hibbins2022phylogenomic,
  title={Phylogenomic approaches to detecting and characterizing introgression},
  author={Hibbins, Mark S and Hahn, Matthew W},
  journal={Genetics},
  volume={220},
  number={2},
  pages={iyab173},
  year={2022},
  publisher={Oxford University Press}
}

@article{bjorner2024detectability,
  title={Detectability of Varied Hybridization Scenarios Using Genome-Scale Hybrid Detection Methods},
  author={Bjorner, Marianne B and Molloy, Erin K and Dewey, Colin N and Solis-Lemus, Claudia},
  journal={Bulletin of the Society of Systematic Biologists},
  volume={3},
  number={1},
  year={2024}
}

@article{frankel2023summary,
  title={Summary tests of introgression are highly sensitive to rate variation across lineages},
  author={Frankel, Lauren E and An{\'e}, C{\'e}cile},
  journal={Systematic Biology},
  volume={72},
  number={6},
  pages={1357--1369},
  year={2023},
  publisher={Oxford University Press US}
}

@article{koppetsch2024towards,
  title={Towards reliable detection of introgression in the presence of among-species rate variation},
  author={Koppetsch, Thore and Malinsky, Milan and Matschiner, Michael},
  journal={Systematic biology},
  volume={73},
  number={5},
  pages={769--788},
  year={2024},
  publisher={Oxford University Press US}
}

@article{pattersonD,
	an = {22960212},
	author = {Patterson, Nick and Moorjani, Priya and Luo, Yontao and Mallick, Swapan and Rohland, Nadin and Zhan, Yiping and Genschoreck, Teri and Webster, Teresa and Reich, David},
	date = {2012/11/},
	db = {PubMed},
	doi = {10.1534/genetics.112.145037},
	et = {2012/09/07},
	isbn = {1943-2631; 0016-6731},
	j2 = {Genetics},
	journal = {Genetics},
	l2 = {https://www.ncbi.nlm.nih.gov/pmc/articles/PMC3522152/},
	la = {eng},
	month = {11},
	number = {3},
	pages = {1065--1093},
	publisher = {Genetics Society of America},
	title = {Ancient admixture in human history},
	ty = {JOUR},
	u1 = {22960212{$[$}pmid{$]$}},
	u2 = {PMC3522152{$[$}pmcid{$]$}},
	u4 = {genetics.112.145037{$[$}PII{$]$}},
	volume = {192},
	year = {2012}
}

@article{allman2022gene,
  title={Gene tree discord, simplex plots, and statistical tests under the coalescent},
  author={Allman, Elizabeth S and Mitchell, Jonathan D and Rhodes, John A},
  journal={Systematic Biology},
  volume={71},
  number={4},
  pages={929--942},
  year={2022},
  publisher={Oxford University Press}
}

@article{rhodes2021mscquartets,
  title={MSCquartets 1.0: quartet methods for species trees and networks under the multispecies coalescent model in R},
  author={Rhodes, John A and Ba{\~n}os, Hector and Mitchell, Jonathan D and Allman, Elizabeth S},
  journal={Bioinformatics},
  volume={37},
  number={12},
  pages={1766--1768},
  year={2021},
  publisher={Oxford University Press}
}

@article{wu2024ultrafast,
  title={Ultrafast learning of four-node hybridization cycles in phylogenetic networks using algebraic invariants},
  author={Wu, Zhaoxing and Sol{\'\i}s-Lemus, Claudia},
  journal={Bioinformatics Advances},
  volume={4},
  number={1},
  pages={vbae014},
  year={2024},
  publisher={Oxford University Press}
}

@article{leache2014influence,
  title={The influence of gene flow on species tree estimation: a simulation study},
  author={Leach{\'e}, Adam D and Harris, Rebecca B and Rannala, Bruce and Yang, Ziheng},
  journal={Systematic biology},
  volume={63},
  number={1},
  pages={17--30},
  year={2014},
  publisher={Oxford University Press}
}

@article{solis2016inconsistency,
  title={Inconsistency of species tree methods under gene flow},
  author={Sol{\'\i}s-Lemus, Claudia and Yang, Mengyao and An{\'e}, C{\'e}cile},
  journal={Systematic biology},
  volume={65},
  number={5},
  pages={843--851},
  year={2016},
  publisher={Oxford University Press}
}

@article{zhu2016light,
  title={In the light of deep coalescence: revisiting trees within networks},
  author={Zhu, Jiafan and Yu, Yun and Nakhleh, Luay},
  journal={BMC bioinformatics},
  volume={17},
  pages={271--282},
  year={2016},
  publisher={Springer}
}

@article{jiao2020impact,
  title={The impact of cross-species gene flow on species tree estimation},
  author={Jiao, Xiyun and Flouri, Tom{\'a}{\v{s}} and Rannala, Bruce and Yang, Ziheng},
  journal={Systematic Biology},
  volume={69},
  number={5},
  pages={830--847},
  year={2020},
  publisher={Oxford University Press}
}

@article{long2018effect,
  title={The effect of gene flow on coalescent-based species-tree inference},
  author={Long, Colby and Kubatko, Laura},
  journal={Systematic biology},
  volume={67},
  number={5},
  pages={770--785},
  year={2018},
  publisher={Oxford University Press}
}

@article{hibbins2019timing,
  title={The timing and direction of introgression under the multispecies network coalescent},
  author={Hibbins, Mark S and Hahn, Matthew W},
  journal={Genetics},
  volume={211},
  number={3},
  pages={1059--1073},
  year={2019},
  publisher={Oxford University Press}
}

@article{peng2024likelihood,
  title={A Likelihood Ratio Test for Hybridization Under the Multispecies Coalescent},
  author={Peng, Jing and Kong, Sungsik and Kubatko, Laura},
  journal={Bulletin of the Society of Systematic Biologists},
  volume={3},
  number={2},
  year={2024}
}

@book{huson2010phylogenetic,
  title={Phylogenetic networks: concepts, algorithms and applications},
  author={Huson, Daniel H and Rupp, Regula and Scornavacca, Celine},
  year={2010},
  publisher={Cambridge University Press}
}

@article{huson2011survey,
  title={A survey of combinatorial methods for phylogenetic networks},
  author={Huson, Daniel H and Scornavacca, Celine},
  journal={Genome biology and evolution},
  volume={3},
  pages={23--35},
  year={2011},
  publisher={Oxford University Press}
}

@article{makarenkov2001t,
  title={T-REX: reconstructing and visualizing phylogenetic trees and reticulation networks},
  author={Makarenkov, Vladimir},
  journal={Bioinformatics},
  volume={17},
  number={7},
  pages={664--668},
  year={2001},
  publisher={Oxford University Press}
}

@article{makarenkov2004phylogenetic,
  title={From a phylogenetic tree to a reticulated network},
  author={Makarenkov, Vladimir and Legendre, Pierre},
  journal={Journal of Computational Biology},
  volume={11},
  number={1},
  pages={195--212},
  year={2004},
  publisher={Mary Ann Liebert, Inc.}
}

@article{makarenkov2004modelling,
  title={Modelling phylogenetic relationships using reticulated networks},
  author={Makarenkov, Vladimir and Legendre, Pierre and Desdevises, Yves},
  journal={Zoologica scripta},
  volume={33},
  number={1},
  pages={89--96},
  year={2004},
  publisher={Wiley Online Library}
}

@article{legendre2002reconstruction,
  title={Reconstruction of biogeographic and evolutionary networks using reticulograms},
  author={Legendre, Pierre and Makarenkov, Vladimir},
  journal={Systematic Biology},
  volume={51},
  number={2},
  pages={199--216},
  year={2002},
  publisher={Society of Systematic Zoology}
}

@article{makarenkov2006phylogenetic,
  title={Phylogenetic network construction approaches},
  author={Makarenkov, Vladimir and Kevorkov, Dmytro and Legendre, Pierre},
  journal={Applied mycology and biotechnology},
  volume={6},
  pages={61--97},
  year={2006},
  publisher={Elsevier}
}

@article{bandelt1992split,
  title={Split decomposition: a new and useful approach to phylogenetic analysis of distance data},
  author={Bandelt, Hans-J{\"u}rgen and Dress, Andreas WM},
  journal={Molecular phylogenetics and evolution},
  volume={1},
  number={3},
  pages={242--252},
  year={1992},
  publisher={Elsevier}
}

@article{bandelt1992canonical,
  title={A canonical decomposition theory for metrics on a finite set},
  author={Bandelt, Hans-J{\"u}rgen and Dress, Andreas WM},
  journal={Advances in mathematics},
  volume={92},
  number={1},
  pages={47--105},
  year={1992},
  publisher={Elsevier}
}

@article{bandelt1995mitochondrial,
  title={Mitochondrial portraits of human populations using median networks.},
  author={Bandelt, Hans-J and Forster, Peter and Sykes, Bryan C and Richards, Martin B},
  journal={Genetics},
  volume={141},
  number={2},
  pages={743--753},
  year={1995},
  publisher={Oxford University Press}
}

@article{bandelt1999median,
  title={Median-joining networks for inferring intraspecific phylogenies.},
  author={Bandelt, Hans-Jurgen and Forster, Peter and R{\"o}hl, Arne},
  journal={Molecular biology and evolution},
  volume={16},
  number={1},
  pages={37--48},
  year={1999}
}

@article{bandelt2000median,
  title={Median networks: speedy construction and greedy reduction, one simulation, and two case studies from human mtDNA},
  author={Bandelt, Hans-J{\"u}rgen and Macaulay, Vincent and Richards, Martin},
  journal={Molecular phylogenetics and evolution},
  volume={16},
  number={1},
  pages={8--28},
  year={2000},
  publisher={Elsevier}
}

@article{huson1998splitstree,
  title={SplitsTree: analyzing and visualizing evolutionary data.},
  author={Huson, Daniel H.},
  journal={Bioinformatics (Oxford, England)},
  volume={14},
  number={1},
  pages={68--73},
  year={1998}
}

@article{huson2006application,
  title={Application of phylogenetic networks in evolutionary studies},
  author={Huson, Daniel H and Bryant, David},
  journal={Molecular biology and evolution},
  volume={23},
  number={2},
  pages={254--267},
  year={2006},
  publisher={Oxford University Press}
}

@article{bryant2004neighbor,
  title={Neighbor-net: an agglomerative method for the construction of phylogenetic networks},
  author={Bryant, David and Moulton, Vincent},
  journal={Molecular biology and evolution},
  volume={21},
  number={2},
  pages={255--265},
  year={2004},
  publisher={Oxford University Press}
}

@article{templeton1992cladistic,
  title={A cladistic analysis of phenotypic associations with haplotypes inferred from restriction endonuclease mapping and DNA sequence data. III. Cladogram estimation.},
  author={Templeton, Alan R and Crandall, Keith A and Sing, Charles F},
  journal={Genetics},
  volume={132},
  number={2},
  pages={619--633},
  year={1992},
  publisher={Oxford University Press}
}

@article{fitch1997networks,
  title={Networks and viral evolution},
  author={Fitch, Walter M},
  journal={Journal of molecular evolution},
  volume={44},
  pages={S65--S75},
  year={1997},
  publisher={Springer}
}

@article{excoffier1994using,
  title={Using allele frequencies and geographic subdivision to reconstruct gene trees within a species: molecular variance parsimony.},
  author={Excoffier, Laurent and Smouse, Peter E},
  journal={Genetics},
  volume={136},
  number={1},
  pages={343--359},
  year={1994},
  publisher={Oxford University Press}
}

@article{paradis2018analysis,
  title={Analysis of haplotype networks: The randomized minimum spanning tree method},
  author={Paradis, Emmanuel},
  journal={Methods in Ecology and Evolution},
  volume={9},
  number={5},
  pages={1308--1317},
  year={2018},
  publisher={Wiley Online Library}
}

@article{leigh2015popart,
  title={POPART: full-feature software for haplotype network construction.},
  author={Leigh, Jessica W and Bryant, David and Nakagawa, Shinichi},
  journal={Methods in Ecology \& Evolution},
  volume={6},
  number={9},
  year={2015}
}

@article{joly2007haplotype,
  title={Haplotype networks can be misleading in the presence of missing data},
  author={Joly, Simon and Stevens, Mark I and Van Vuuren, Bettine Jansen},
  journal={Systematic biology},
  volume={56},
  number={5},
  pages={857--862},
  year={2007},
  publisher={Oxford University Press}
}

@article{gambette2012quartets,
  title={Quartets and unrooted phylogenetic networks},
  author={Gambette, Philippe and Berry, Vincent and Paul, Christophe},
  journal={Journal of bioinformatics and computational biology},
  volume={10},
  number={04},
  pages={1250004},
  year={2012},
  publisher={World Scientific}
}

@article{huber2013encoding,
  title={Encoding and constructing 1-nested phylogenetic networks with trinets},
  author={Huber, Katharina T and Moulton, Vincent},
  journal={Algorithmica},
  volume={66},
  pages={714--738},
  year={2013},
  publisher={Springer}
}

@article{pickrell2012inference,
  title={Inference of population splits and mixtures from genome-wide allele frequency data},
  author={Pickrell, Joseph and Pritchard, Jonathan},
  journal={Nature Precedings},
  pages={1--1},
  year={2012},
  publisher={Nature Publishing Group UK London}
}

@article{grunewald2007qnet,
  title={QNet: an agglomerative method for the construction of phylogenetic networks from weighted quartets},
  author={Gr{\"u}newald, Stefan and Forslund, Kristoffer and Dress, Andreas and Moulton, Vincent},
  journal={Molecular Biology and Evolution},
  volume={24},
  number={2},
  pages={532--538},
  year={2007},
  publisher={Oxford University Press}
}

@article{yang2013quartet,
  title={Quartet-net: a quartet-based method to reconstruct phylogenetic networks},
  author={Yang, Jialiang and Gr{\"u}newald, Stefan and Wan, Xiu-Feng},
  journal={Molecular biology and evolution},
  volume={30},
  number={5},
  pages={1206--1217},
  year={2013},
  publisher={Oxford University Press}
}

@article{willems2014new,
  title={A new efficient algorithm for inferring explicit hybridization networks following the Neighbor-Joining principle},
  author={Willems, Matthieu and Tahiri, Nadia and Makarenkov, Vladimir},
  journal={Journal of bioinformatics and computational biology},
  volume={12},
  number={05},
  pages={1450024},
  year={2014},
  publisher={World Scientific}
}

@article{huson2012dendroscope,
  title={Dendroscope 3: an interactive tool for rooted phylogenetic trees and networks},
  author={Huson, Daniel H and Scornavacca, Celine},
  journal={Systematic biology},
  volume={61},
  number={6},
  pages={1061--1067},
  year={2012},
  publisher={Oxford University Press}
}

@article{wang2013bimlr,
  title={BIMLR: A method for constructing rooted phylogenetic networks from rooted phylogenetic trees},
  author={Wang, Juan and Guo, Maozu and Xing, Linlin and Che, Kai and Liu, Xiaoyan and Wang, Chunyu},
  journal={Gene},
  volume={527},
  number={1},
  pages={344--351},
  year={2013},
  publisher={Elsevier}
}

@article{fogg2023phylocoalsimulations,
  title={PhyloCoalSimulations: a simulator for network multispecies coalescent models, including a new extension for the inheritance of gene flow},
  author={Fogg, John and Allman, Elizabeth S and An{\'e}, C{\'e}cile},
  journal={Systematic Biology},
  volume={72},
  number={5},
  pages={1171--1179},
  year={2023},
  publisher={Oxford University Press US}
}

@article{lutteropp2022netrax,
  title={NetRAX: accurate and fast maximum likelihood phylogenetic network inference},
  author={Lutteropp, Sarah and Scornavacca, C{\'e}line and Kozlov, Alexey M and Morel, Benoit and Stamatakis, Alexandros},
  journal={Bioinformatics},
  volume={38},
  number={15},
  pages={3725--3733},
  year={2022},
  publisher={Oxford University Press}
}

@article{markin2022rf,
  title={RF-Net 2: Fast inference of virus reassortment and hybridization networks},
  author={Markin, Alexey and Wagle, Sanket and Anderson, Tavis K and Eulenstein, Oliver},
  journal={Bioinformatics},
  volume={38},
  number={8},
  pages={2144--2152},
  year={2022},
  publisher={Oxford University Press}
}

@article{jones2013statistical,
  title={Statistical inference of allopolyploid species networks in the presence of incomplete lineage sorting},
  author={Jones, Graham and Sagitov, Serik and Oxelman, Bengt},
  journal={Systematic biology},
  volume={62},
  number={3},
  pages={467--478},
  year={2013},
  publisher={Oxford University Press}
}

@article{holtgrefe2025squirrel,
  title={Squirrel: Reconstructing semi-directed phylogenetic level-1 networks from four-leaved networks or sequence alignments},
  author={Holtgrefe, Niels and Huber, Katharina T and van Iersel, Leo and Jones, Mark and Martin, Samuel and Moulton, Vincent},
  journal={Molecular Biology and Evolution},
  volume={42},
  number={4},
  pages={msaf067},
  year={2025},
  publisher={Oxford University Press UK}
}

@article{allman2019nanuq,
  title={NANUQ: a method for inferring species networks from gene trees under the coalescent model},
  author={Allman, Elizabeth S and Ba{\~n}os, Hector and Rhodes, John A},
  journal={Algorithms for Molecular Biology},
  volume={14},
  pages={1--25},
  year={2019},
  publisher={Springer}
}

@article{forster2020phylogenetic,
  title={Phylogenetic network analysis of SARS-CoV-2 genomes},
  author={Forster, Peter and Forster, Lucy and Renfrew, Colin and Forster, Michael},
  journal={Proceedings of the National Academy of Sciences},
  volume={117},
  number={17},
  pages={9241--9243},
  year={2020},
  publisher={National Academy of Sciences}
}

@article{sanchez2020median,
  title={Median-joining network analysis of SARS-CoV-2 genomes is neither phylogenetic nor evolutionary},
  author={S{\'a}nchez-Pacheco, Santiago J and Kong, Sungsik and Pulido-Santacruz, Paola and Murphy, Robert W and Kubatko, Laura},
  journal={Proceedings of the National Academy of Sciences},
  volume={117},
  number={23},
  pages={12518--12519},
  year={2020},
  publisher={National Academy of Sciences}
}

@article{mavian2020sampling,
  title={Sampling bias and incorrect rooting make phylogenetic network tracing of SARS-COV-2 infections unreliable},
  author={Mavian, Carla and Pond, Sergei Kosakovsky and Marini, Simone and Magalis, Brittany Rife and Vandamme, Anne-Mieke and Dellicour, Simon and Scarpino, Samuel V and Houldcroft, Charlotte and Villabona-Arenas, Julian and Paisie, Taylor K and others},
  journal={Proceedings of the National Academy of Sciences},
  volume={117},
  number={23},
  pages={12522--12523},
  year={2020},
  publisher={National Academy of Sciences}
}

@article{chookajorn2020evolving,
  title={Evolving COVID-19 conundrum and its impact},
  author={Chookajorn, Thanat},
  journal={Proceedings of the National Academy of Sciences},
  volume={117},
  number={23},
  pages={12520--12521},
  year={2020},
  publisher={National Academy of Sciences}
}

@article{bryant2023neighbornet,
  title={NeighborNet: improved algorithms and implementation},
  author={Bryant, David and Huson, Daniel H},
  journal={Frontiers in bioinformatics},
  volume={3},
  pages={1178600},
  year={2023},
  publisher={Frontiers Media SA}
}

@article{francis2018tree,
  title={Tree-based unrooted phylogenetic networks},
  author={Francis, Andrew and Huber, Katharina T and Moulton, Vincent},
  journal={Bulletin of mathematical biology},
  volume={80},
  number={2},
  pages={404--416},
  year={2018},
  publisher={Springer}
}

@article{gambette2017uprooted,
  title={Uprooted phylogenetic networks},
  author={Gambette, Philippe and Huber, Katharina T and Scholz, Guillaume E},
  journal={Bulletin of mathematical biology},
  volume={79},
  number={9},
  pages={2022--2048},
  year={2017},
  publisher={Springer}
}

@article{mo2024applications,
  title={Applications of machine learning in phylogenetics},
  author={Mo, Yu K and Hahn, Matthew W and Smith, Megan L},
  journal={Molecular Phylogenetics and Evolution},
  volume={196},
  pages={108066},
  year={2024},
  publisher={Elsevier}
}

@article{van2014trinets,
  title={Trinets encode tree-child and level-2 phylogenetic networks},
  author={Van Iersel, Leo and Moulton, Vincent},
  journal={Journal of mathematical biology},
  volume={68},
  number={7},
  pages={1707--1729},
  year={2014},
  publisher={Springer}
}

@article{huber2010practical,
  title={A practical algorithm for reconstructing level-1 phylogenetic networks},
  author={Huber, Katharina T and Van Iersel, Leo and Kelk, Steven and Suchecki, Radoslaw},
  journal={IEEE/ACM Transactions on Computational Biology and Bioinformatics},
  volume={8},
  number={3},
  pages={635--649},
  year={2010},
  publisher={IEEE}
}

@article{oldman2016trilonet,
  title={Trilonet: piecing together small networks to reconstruct reticulate evolutionary histories},
  author={Oldman, James and Wu, Taoyang and Van Iersel, Leo and Moulton, Vincent},
  journal={Molecular biology and evolution},
  volume={33},
  number={8},
  pages={2151--2162},
  year={2016},
  publisher={Oxford University Press}
}

@article{van2009uniqueness,
  title={Uniqueness, intractability and exact algorithms: reflections on level-k phylogenetic networks},
  author={Van Iersel, Leo and Kelk, Steven and Mnich, Matthias},
  journal={Journal of Bioinformatics and Computational Biology},
  volume={7},
  number={04},
  pages={597--623},
  year={2009},
  publisher={World Scientific}
}

@article{poormohammadi2014tripnet,
  title={TripNet: a method for constructing rooted phylogenetic networks from rooted triplets},
  author={Poormohammadi, Hadi and Eslahchi, Changiz and Tusserkani, Ruzbeh},
  journal={PloS one},
  volume={9},
  number={9},
  pages={e106531},
  year={2014},
  publisher={Public Library of Science San Francisco, USA}
}

@article{erdHos2019class,
  title={A class of phylogenetic networks reconstructable from ancestral profiles},
  author={Erd{\H{o}}s, P{\'e}ter L and Semple, Charles and Steel, Mike},
  journal={Mathematical biosciences},
  volume={313},
  pages={33--40},
  year={2019},
  publisher={Elsevier}
}

@article{jansson2006algorithms,
  title={Algorithms for combining rooted triplets into a galled phylogenetic network},
  author={Jansson, Jesper and Nguyen, Nguyen Bao and Sung, Wing-Kin},
  journal={SIAM Journal on Computing},
  volume={35},
  number={5},
  pages={1098--1121},
  year={2006},
  publisher={SIAM}
}

@article{van2009constructing,
  title={Constructing level-2 phylogenetic networks from triplets},
  author={Van Iersel, Leo and Keijsper, Judith and Kelk, Steven and Stougie, Leen and Hagen, Ferry and Boekhout, Teun},
  journal={IEEE/ACM Transactions on Computational Biology and Bioinformatics},
  volume={6},
  number={4},
  pages={667--681},
  year={2009},
  publisher={IEEE}
}

@inproceedings{gambette2009structure,
  title={The structure of level-k phylogenetic networks},
  author={Gambette, Philippe and Berry, Vincent and Paul, Christophe},
  booktitle={Annual Symposium on Combinatorial Pattern Matching},
  pages={289--300},
  year={2009},
  organization={Springer}
}

@article{van2017binets,
  title={Binets: fundamental building blocks for phylogenetic networks},
  author={Van Iersel, Leo and Moulton, Vincent and de Swart, Eveline and Wu, Taoyang},
  journal={Bulletin of mathematical biology},
  volume={79},
  number={5},
  pages={1135--1154},
  year={2017},
  publisher={Springer}
}

@article{linz2020caterpillars,
  title={Caterpillars on three and four leaves are sufficient to reconstruct binary normal networks},
  author={Linz, Simone and Semple, Charles},
  journal={Journal of Mathematical Biology},
  volume={81},
  number={4},
  pages={961--980},
  year={2020},
  publisher={Springer}
}

@article{van2018leaf,
  title={Leaf-reconstructibility of phylogenetic networks},
  author={Van Iersel, Leo and Moulton, Vincent},
  journal={SIAM Journal on Discrete Mathematics},
  volume={32},
  number={3},
  pages={2047--2066},
  year={2018},
  publisher={SIAM}
}

@article{keijsper2014reconstructing,
  title={Reconstructing a phylogenetic level-1 network from quartets},
  author={Keijsper, JCM and Pendavingh, RA},
  journal={Bulletin of Mathematical Biology},
  volume={76},
  number={10},
  pages={2517--2541},
  year={2014},
  publisher={Springer}
}

@article{huber2018quarnet,
  title={Quarnet inference rules for level-1 networks},
  author={Huber, Katharina T and Moulton, Vincent and Semple, Charles and Wu, Taoyang},
  journal={Bulletin of mathematical biology},
  volume={80},
  number={8},
  pages={2137--2153},
  year={2018},
  publisher={Springer}
}

@article{reyhani2017rpnch,
  title={RPNCH: A method for constructing rooted phylogenetic networks from rooted triplets based on height function},
  author={Reyhani, Mohammad Hossein and Poormohammadi, Hadi},
  journal={Archives of Advances in Biosciences},
  volume={8},
  number={4},
  pages={14--20},
  year={2017}
}

@article{kelk2011elusiveness,
  title={On the elusiveness of clusters},
  author={Kelk, Steven and Scornavacca, Celine and Van Iersel, Leo},
  journal={IEEE/ACM Transactions on Computational Biology and Bioinformatics},
  volume={9},
  number={2},
  pages={517--534},
  year={2011},
  publisher={IEEE}
}

@article{gambette2012encodings,
  title={On encodings of phylogenetic networks of bounded level},
  author={Gambette, Philippe and Huber, Katharina T},
  journal={Journal of mathematical biology},
  volume={65},
  number={1},
  pages={157--180},
  year={2012},
  publisher={Springer}
}

@article{fischer2020space,
  title={The space of tree-based phylogenetic networks},
  author={Fischer, Mareike and Francis, Andrew},
  journal={Bulletin of Mathematical Biology},
  volume={82},
  number={6},
  pages={70},
  year={2020},
  publisher={Springer}
}

@article{fischer2021unrooted,
  title={Unrooted non-binary tree-based phylogenetic networks},
  author={Fischer, Mareike and Herbst, Lina and Galla, Michelle and Long, Yangjing and Wicke, Kristina},
  journal={Discrete applied mathematics},
  volume={294},
  pages={10--30},
  year={2021},
  publisher={Elsevier}
}

@article{cardona2020counting,
  title={Counting and enumerating tree-child networks and their subclasses},
  author={Cardona, Gabriel and Zhang, Louxin},
  journal={Journal of Computer and System Sciences},
  volume={114},
  pages={84--104},
  year={2020},
  publisher={Elsevier}
}

@article{huebler2019constructing,
  title={Constructing semi-directed level-1 phylogenetic networks from quarnets},
  author={Huebler, Sophia and Morris, Rachel and Rusinko, Joseph and Tao, Yifei},
  journal={arXiv preprint arXiv:1910.00048},
  year={2019}
}

@article{van2010phylogenetic,
  title={Phylogenetic networks do not need to be complex: using fewer reticulations to represent conflicting clusters},
  author={Van Iersel, Leo and Kelk, Steven and Rupp, Regula and Huson, Daniel},
  journal={Bioinformatics},
  volume={26},
  number={12},
  pages={i124--i131},
  year={2010},
  publisher={Oxford University Press}
}

@article{willems2018building,
  title={Building explicit hybridization networks using the maximum likelihood and Neighbor-Joining approaches},
  author={Willems, Matthieu and Tahiri, Nadia and Makarenkov, Vladimir},
  journal={Archives of Data Science, Series A (Online First)},
  volume={4},
  number={1},
  pages={04},
  year={2018}
}

@article{wang2013lnetwork,
  title={Lnetwork: an efficient and effective method for constructing phylogenetic networks},
  author={Wang, Juan and Guo, Maozu and Liu, Xiaoyan and Liu, Yang and Wang, Chunyu and Xing, Linlin and Che, Kai},
  journal={Bioinformatics},
  volume={29},
  number={18},
  pages={2269--2276},
  year={2013},
  publisher={Oxford University Press}
}

@inproceedings{linder2004network,
  title={Network (reticulate) evolution: biology, models, and algorithms},
  author={Linder, C Randal and Moret, Bernard ME and Nakhleh, Luay and Warnow, Tandy},
  booktitle={The Ninth Pacific Symposium on Biocomputing (PSB)},
  year={2004}
}

@article{morrison2005networks,
  title={Networks in phylogenetic analysis: new tools for population biology},
  author={Morrison, David A},
  journal={International journal for parasitology},
  volume={35},
  number={5},
  pages={567--582},
  year={2005},
  publisher={Elsevier}
}

@article{bapteste2013networks,
  title={Networks: expanding evolutionary thinking},
  author={Bapteste, Eric and van Iersel, Leo and Janke, Axel and Kelchner, Scot and Kelk, Steven and McInerney, James O and Morrison, David A and Nakhleh, Luay and Steel, Mike and Stougie, Leen and others},
  journal={Trends in Genetics},
  volume={29},
  number={8},
  pages={439--441},
  year={2013},
  publisher={Elsevier}
}

@article{rannala2003bayes,
  title={Bayes estimation of species divergence times and ancestral population sizes using DNA sequences from multiple loci},
  author={Rannala, Bruce and Yang, Ziheng},
  journal={Genetics},
  volume={164},
  number={4},
  pages={1645--1656},
  year={2003},
  publisher={Oxford University Press}
}

@incollection{rannala2020multispecies,
  title={The multispecies coalescent model and species tree inference},
  author={Rannala, Bruce and Leache, A and Edwards, S and Yang, Ziheng},
  year={2020},
  publisher={In Scornavacca, C., Delsuc, F., and Galtier, N., editors, Phylogenetics in the Genomic Era}
}

@article{zhang2018bayesian,
  title={Bayesian inference of species networks from multilocus sequence data},
  author={Zhang, Chi and Ogilvie, Huw A and Drummond, Alexei J and Stadler, Tanja},
  journal={Molecular biology and evolution},
  volume={35},
  number={2},
  pages={504--517},
  year={2018},
  publisher={Oxford University Press}
}

@article{wen2016bayesian,
  title={Bayesian inference of reticulate phylogenies under the multispecies network coalescent},
  author={Wen, Dingqiao and Yu, Yun and Nakhleh, Luay},
  journal={PLoS genetics},
  volume={12},
  number={5},
  pages={e1006006},
  year={2016},
  publisher={Public Library of Science San Francisco, CA USA}
}

@article{gambette2017rearrangement,
  title={Rearrangement moves on rooted phylogenetic networks},
  author={Gambette, Philippe and Van Iersel, Leo and Jones, Mark and Lafond, Manuel and Pardi, Fabio and Scornavacca, Celine},
  journal={PLoS computational biology},
  volume={13},
  number={8},
  pages={e1005611},
  year={2017},
  publisher={Public Library of Science San Francisco, CA USA}
}

@article{huber2016spaces,
  title={Spaces of phylogenetic networks from generalized nearest-neighbor interchange operations},
  author={Huber, Katharina T and Linz, Simone and Moulton, Vincent and Wu, Taoyang},
  journal={Journal of Mathematical Biology},
  volume={72},
  number={3},
  pages={699--725},
  year={2016},
  publisher={Springer}
}

@article{janssen2018exploring,
  title={Exploring the tiers of rooted phylogenetic network space using tail moves},
  author={Janssen, Remie and Jones, Mark and Erd{\H{o}}s, P{\'e}ter L and Van Iersel, Leo and Scornavacca, Celine},
  journal={Bulletin of mathematical biology},
  volume={80},
  number={8},
  pages={2177--2208},
  year={2018},
  publisher={Springer}
}

@article{bordewich2017lost,
  title={Lost in space? Generalising subtree prune and regraft to spaces of phylogenetic networks},
  author={Bordewich, Magnus and Linz, Simone and Semple, Charles},
  journal={Journal of theoretical biology},
  volume={423},
  pages={1--12},
  year={2017},
  publisher={Elsevier}
}

@article{janssen2018heading,
  title={Heading in the right direction? Using head moves to traverse phylogenetic network space},
  author={Janssen, Remie},
  journal={arXiv preprint arXiv:1810.09795},
  year={2018}
}

@article{linz2023exploring,
  title={Exploring spaces of semi-directed level-1 networks},
  author={Linz, Simone and Wicke, Kristina},
  journal={Journal of Mathematical Biology},
  volume={87},
  number={5},
  pages={70},
  year={2023},
  publisher={Springer}
}

@article{degnan2018modeling,
  title={Modeling hybridization under the network multispecies coalescent},
  author={Degnan, James H},
  journal={Systematic biology},
  volume={67},
  number={5},
  pages={786--799},
  year={2018},
  publisher={Oxford University Press}
}

@ARTICLE{Solis-Lemus2016-ux,
  title    = "Inferring Phylogenetic Networks with Maximum Pseudolikelihood
              under Incomplete Lineage Sorting",
  author   = "Sol{\'\i}s-Lemus, Claudia and An{\'e}, C{\'e}cile",
  journal  = "PLoS Genet.",
  volume   =  12,
  number   =  3,
  pages    = "e1005896",
  month    =  mar,
  year     =  2016,
}

@article{wen2018inferring,
  title={Inferring phylogenetic networks using PhyloNet},
  author={Wen, Dingqiao and Yu, Yun and Zhu, Jiafan and Nakhleh, Luay},
  journal={Systematic biology},
  volume={67},
  number={4},
  pages={735--740},
  year={2018},
  publisher={Oxford University Press}
}

@article{yu2015maximum,
  title={A maximum pseudo-likelihood approach for phylogenetic networks},
  author={Yu, Yun and Nakhleh, Luay},
  journal={BMC genomics},
  volume={16},
  year = {2015},
  number={Suppl 10},
  pages={S10},
  publisher={Springer}
}

@inproceedings{yu2015distance,
  title={A distance-based method for inferring phylogenetic networks in the presence of incomplete lineage sorting},
  author={Yu, Yun and Nakhleh, Luay},
  booktitle={International Symposium on Bioinformatics Research and Applications},
  pages={378--389},
  year={2015},
  organization={Springer}
}

@article{zhu2018bayesian,
  title={Bayesian inference of phylogenetic networks from bi-allelic genetic markers},
  author={Zhu, Jiafan and Wen, Dingqiao and Yu, Yun and Meudt, Heidi M and Nakhleh, Luay},
  journal={PLoS computational biology},
  volume={14},
  number={1},
  pages={e1005932},
  year={2018},
  publisher={Public Library of Science San Francisco, CA USA}
}

@article{zhu2018inference,
  title={Inference of species phylogenies from bi-allelic markers using pseudo-likelihood},
  author={Zhu, Jiafan and Nakhleh, Luay},
  journal={Bioinformatics},
  volume={34},
  number={13},
  pages={i376--i385},
  year={2018},
  publisher={Oxford University Press}
}

@article{kong2025inference,
  title={Inference of phylogenetic networks from sequence data using composite likelihood},
  author={Kong, Sungsik and Swofford, David L and Kubatko, Laura S},
  journal={Systematic Biology},
  volume={74},
  number={1},
  pages={53--69},
  year={2025},
  publisher={Oxford University Press US}
}

@article{yang2015bpp,
  title={The BPP program for species tree estimation and species delimitation},
  author={Yang, Ziheng},
  journal={Current Zoology},
  volume={61},
  number={5},
  pages={854--865},
  year={2015},
  publisher={Oxford University Press Oxford, UK}
}

@article{hey2004multilocus,
  title={Multilocus methods for estimating population sizes, migration rates and divergence time, with applications to the divergence of Drosophila pseudoobscura and D. persimilis},
  author={Hey, Jody and Nielsen, Rasmus},
  journal={Genetics},
  volume={167},
  number={2},
  pages={747--760},
  year={2004},
  publisher={Oxford University Press}
}

@article{flouri2020bayesian,
  title={A Bayesian implementation of the multispecies coalescent model with introgression for phylogenomic analysis},
  author={Flouri, Tom{\'a}{\v{s}} and Jiao, Xiyun and Rannala, Bruce and Yang, Ziheng},
  journal={Molecular biology and evolution},
  volume={37},
  number={4},
  pages={1211--1223},
  year={2020},
  publisher={Oxford University Press}
}

@article{flouri2023efficient,
  title={Efficient Bayesian inference under the multispecies coalescent with migration},
  author={Flouri, Tom{\'a}{\v{s}} and Jiao, Xiyun and Huang, Jun and Rannala, Bruce and Yang, Ziheng},
  journal={Proceedings of the National Academy of Sciences},
  volume={120},
  number={44},
  pages={e2310708120},
  year={2023},
  publisher={National Academy of Sciences}
}

@article{jiao2021multispecies,
  title={Multispecies coalescent and its applications to infer species phylogenies and cross-species gene flow},
  author={Jiao, Xiyun and Flouri, Tom{\'a}{\v{s}} and Yang, Ziheng},
  journal={National Science Review},
  volume={8},
  number={12},
  pages={nwab127},
  year={2021},
  publisher={Oxford University Press}
}

@article{rabier2021inference,
  title={On the inference of complex phylogenetic networks by Markov Chain Monte-Carlo},
  author={Rabier, Charles-Elie and Berry, Vincent and Stoltz, Marnus and Santos, Joao D and Wang, Wensheng and Glaszmann, Jean-Christophe and Pardi, Fabio and Scornavacca, Celine},
  journal={PLoS Computational Biology},
  volume={17},
  number={9},
  pages={e1008380},
  year={2021},
  publisher={Public Library of Science San Francisco, CA USA}
}

@article{huber2017reconstructing,
  title={Reconstructing phylogenetic level-1 networks from nondense binet and trinet sets},
  author={Huber, Katharina T and Van Iersel, Leo and Moulton, Vincent and Scornavacca, Celine and Wu, Taoyang},
  journal={Algorithmica},
  volume={77},
  number={1},
  pages={173--200},
  year={2017},
  publisher={Springer}
}

@article{jones2017,
  title = {Bayesian Phylogenetic Analysis for Diploid and Allotetraploid Species Networks},
  author = {Jones, Graham},
  year = {2017},
  month = apr,
  doi = {10.1101/129361},
  urldate = {2024-05-09},
  langid = {english},
  journal = {bioRxiv}
}

@article{one2019one,
  title={One thousand plant transcriptomes and the phylogenomics of green plants},
  author = {1000Plant-Transcriptomes-Initiative},
  journal={Nature},
  volume={574},
  number={7780},
  pages={679--685},
  year={2019},
  publisher={Nature Publishing Group UK London}
}

@inproceedings{hejase2018fastnet,
  title={Fastnet: fast and accurate statistical inference of phylogenetic networks using large-scale genomic sequence data},
  author={Hejase, Hussein A and VandePol, Natalie and Bonito, Gregory M and Liu, Kevin J},
  booktitle={RECOMB International conference on Comparative Genomics},
  pages={242--259},
  year={2018},
  organization={Springer}
}

@article{kolbow2025method,
  title={A method for massively scalable phylogenetic network inference},
  author={Kolbow, Nathan and Kong, Sungsik and Solis-Lemus, Claudia},
  journal={bioRxiv},
  pages={2025--05},
  year={2025},
  publisher={Cold Spring Harbor Laboratory}
}

@article{allman2024tinnik,
  title={TINNiK: inference of the tree of blobs of a species network under the coalescent model},
  author={Allman, Elizabeth S and Ba{\~n}os, Hector and Mitchell, Jonathan D and Rhodes, John A},
  journal={Algorithms for Molecular Biology},
  volume={19},
  number={1},
  pages={23},
  year={2024},
  publisher={Springer}
}

@article{allman2024nanuq,
  title={NANUQ+: a divide-and-conquer approach to network estimation},
  author={Allman, Elizabeth S and Ba{\~n}os, Hector and Rhodes, John A and Wicke, Kristina},
  journal={bioRxiv},
  pages={2024--10},
  year={2024},
  publisher={Cold Spring Harbor Laboratory}
}

@article{allman2023tree,
  title={The tree of blobs of a species network: identifiability under the coalescent},
  author={Allman, Elizabeth S and Ba{\~n}os, Hector and Mitchell, Jonathan D and Rhodes, John A},
  journal={Journal of mathematical biology},
  volume={86},
  number={1},
  pages={10},
  year={2023},
  publisher={Springer}
}

@article{warnow2025advances,
  title={Advances in estimating level-1 phylogenetic networks from unrooted SNPs},
  author={Warnow, Tandy and Tabatabaee, Yasamin and Evans, Steven N},
  journal={Journal of Computational Biology},
  volume={32},
  number={1},
  pages={3--27},
  year={2025},
  publisher={Mary Ann Liebert, Inc., publishers 140 Huguenot Street, 3rd Floor New~…}
}

@article{huber2024quarnets,
  title={When are quarnets sufficient to reconstruct semi-directed phylogenetic networks?},
  author={Huber, Katharina T and van Iersel, Leo and Jones, Mark and Moulton, Vincent and Nipius, Leonie Veenema},
  journal={arXiv preprint arXiv:2408.12997},
  year={2024}
}

@article{Barton2022,
  doi = {10.48550/ARXIV.2211.11919},  
  author = {Barton, Travis and Gross, Elizabeth and Long, Colby and Rusinko, Joseph},
  title = {Statistical learning with phylogenetic network invariants},
  journal = {arXiv},
  year = {2022},
  copyright = {arXiv.org perpetual, non-exclusive license}
}

@article{Gross2018,
author = {Gross, Elizabeth and Long, Colby},
title = {Distinguishing Phylogenetic Networks},
journal = {SIAM Journal on Applied Algebra and Geometry},
volume = {2},
number = {1},
pages = {72-93},
year = {2018},
doi = {10.1137/17M1134238},
URL = {https://doi.org/10.1137/17M1134238},
eprint = {https://doi.org/10.1137/17M1134238}}

@article{gross2021distinguishing,
  title={Distinguishing level-1 phylogenetic networks on the basis of data generated by Markov processes},
  author={Gross, Elizabeth and van Iersel, Leo and Janssen, Remie and Jones, Mark and Long, Colby and Murakami, Yukihiro},
  journal={Journal of Mathematical Biology},
  volume={83},
  number={3},
  pages={1--24},
  year={2021},
  publisher={Springer}
}

@article{ardiyansyah2021distinguishing,
  title={Distinguishing Level-2 Phylogenetic Networks Using Phylogenetic Invariants},
  author={Ardiyansyah, Muhammad},
  journal={arXiv preprint arXiv:2104.12479},
  year={2021}
}

@article{englander2025identifiability,
  title={Identifiability of phylogenetic level-2 networks under the jukes-cantor model},
  author={Englander, Aviva K and Frohn, Martin and Gross, Elizabeth and Holtgrefe, Niels and van Iersel, Leo and Jones, Mark and Sullivant, Seth},
  journal={bioRxiv},
  pages={2025--04},
  year={2025},
  publisher={Cold Spring Harbor Laboratory}
}

@article{cummings2024invariants,
  title={Invariants for level-1 phylogenetic networks under the Cavendar-Farris-Neyman model},
  author={Cummings, Joseph and Hollering, Benjamin and Manon, Christopher},
  journal={Advances in Applied Mathematics},
  volume={153},
  pages={102633},
  year={2024},
  publisher={Elsevier}
}

@article{martin2023algebraic,
  title={Algebraic invariants for inferring 4-leaf semi-directed phylogenetic networks},
  author={Martin, Samuel and Holtgrefe, Niels and Moulton, Vincent and Leggett, Richard M},
  journal={bioRxiv},
  pages={2023--09},
  year={2023},
  publisher={Cold Spring Harbor Laboratory}
}

@article{baum2007concordance,
  title={Concordance trees, concordance factors, and the exploration of reticulate genealogy},
  author={Baum, David A},
  journal={Taxon},
  volume={56},
  number={2},
  pages={417--426},
  year={2007},
  publisher={Wiley Online Library}
}

@article{banos2019identifying,
  title={Identifying species network features from gene tree quartets under the coalescent model},
  author={Ba{\~n}os, Hector},
  journal={Bulletin of mathematical biology},
  volume={81},
  number={2},
  pages={494--534},
  year={2019},
  publisher={Springer}
}

@article{pardi2015reconstructible,
  title={Reconstructible phylogenetic networks: do not distinguish the indistinguishable},
  author={Pardi, Fabio and Scornavacca, Celine},
  journal={PLoS computational biology},
  volume={11},
  number={4},
  pages={e1004135},
  year={2015},
  publisher={Public Library of Science San Francisco, CA USA}
}

@article{zhu2017displayed,
  title={Displayed trees do not determine distinguishability under the network multispecies coalescent},
  author={Zhu, Sha and Degnan, James H},
  journal={Systematic biology},
  volume={66},
  number={2},
  pages={283--298},
  year={2017},
  publisher={Oxford University Press}
}

@article{allman2022identifiability,
  title={Identifiability of species network topologies from genomic sequences using the logdet distance},
  author={Allman, Elizabeth S and Ba{\~n}os, Hector and Rhodes, John A},
  journal={Journal of mathematical biology},
  volume={84},
  number={5},
  pages={35},
  year={2022},
  publisher={Springer}
}

@article{ane2024anomalous,
  title={Anomalous networks under the multispecies coalescent: theory and prevalence},
  author={An{\'e}, C{\'e}cile and Fogg, John and Allman, Elizabeth S and Ba{\~n}os, Hector and Rhodes, John A},
  journal={Journal of Mathematical Biology},
  volume={88},
  number={3},
  pages={29},
  year={2024},
  publisher={Springer}
}

@article{allman2024identifiability,
  title={Identifiability of level-1 species networks from gene tree quartets},
  author={Allman, Elizabeth S and Ba{\~n}os, Hector and Garrote-Lopez, Marina and Rhodes, John A},
  journal={Bulletin of Mathematical Biology},
  volume={86},
  number={9},
  pages={110},
  year={2024},
  publisher={Springer}
}

@article{allman2025beyond,
  title={Beyond level-1: Identifiability of a class of galled tree-child networks},
  author={Allman, Elizabeth S and Ane, Cecile and Banos, Hector and Rhodes, John A},
  journal={arXiv preprint arXiv:2504.21116},
  year={2025}
}

@article{rhodes2025identifying,
  title={Identifying circular orders for blobs in phylogenetic networks},
  author={Rhodes, John A and Banos, Hector and Xu, Jingcheng and An{\'e}, C{\'e}cile},
  journal={Advances in Applied Mathematics},
  volume={163},
  pages={102804},
  year={2025},
  publisher={Elsevier}
}

@article{xu2023identifiability,
  title={Identifiability of local and global features of phylogenetic networks from average distances},
  author={Xu, Jingcheng and An{\'e}, C{\'e}cile},
  journal={Journal of Mathematical Biology},
  volume={86},
  number={1},
  pages={12},
  year={2023},
  publisher={Springer}
}

@article{frohn2024invariants,
  title={Invariants for level-1 phylogenetic networks under the random walk 4-state Markov model},
  author={Frohn, Martin and Holtgrefe, Niels and van Iersel, Leo and Jones, Mark and Kelk, Steven},
  journal={arXiv preprint arXiv:2407.11720},
  year={2024}
}

@article{frohn2025reconstructing,
  title={Reconstructing semi-directed level-1 networks using few quarnets},
  author={Frohn, Martin and Holtgrefe, Niels and van Iersel, Leo and Jones, Mark and Kelk, Steven},
  journal={Journal of Computer and System Sciences},
  volume={152},
  pages={103655},
  year={2025},
  publisher={Elsevier}
}

@article{maxfield2025dissimilarity,
  title={A dissimilarity measure for semidirected networks},
  author={Maxfield, Michael and Xu, Jingcheng and An{\'e}, C{\'e}cile},
  journal={IEEE Transactions on Computational Biology and Bioinformatics},
  year={2025},
  publisher={IEEE}
}

@article{dinh2025misspecification,
  title={Misspecification Strikes: ASTRAL can Mislead in the Presence of Hybridization, even for Nonanomalous Scenarios},
  author={Dinh, Vu and Ba{\~n}os, Hector},
  journal={Molecular Biology and Evolution},
  volume={42},
  number={3},
  pages={msaf049},
  year={2025},
  publisher={Oxford University Press UK}
}

@article{van2022algorithm,
  title={An algorithm for reconstructing level-2 phylogenetic networks from trinets},
  author={van Iersel, Leo and Kole, Sjors and Moulton, Vincent and Nipius, Leonie},
  journal={Information Processing Letters},
  volume={178},
  pages={106300},
  year={2022},
  publisher={Elsevier}
}

@article{huber2023computing,
  title={Computing consensus networks for collections of 1-nested phylogenetic networks},
  author={Huber, Katharina T and Moulton, Vincent and Spillner, Andreas},
  journal={Journal of Graph Algorithms and Applications},
  volume={27},
  number={7},
  pages={541--563},
  year={2023},
  publisher={Brown University}
}

@article{moulton2017cubic,
  title={A cubic-time algorithm for computing the trinet distance between level-1 networks},
  author={Moulton, Vincent and Oldman, James and Wu, Taoyang},
  journal={Information Processing Letters},
  volume={123},
  pages={36--41},
  year={2017},
  publisher={Elsevier}
}

@article{tamiji2025impact,
  title={Impact of Data Error on Phylogenetic Network Inference from Gene Trees Under the Multispecies Network Coalescent},
  author={Tamiji, Mehrdad and Sapoval, Nicolae and Nakhleh, Luay},
  journal={bioRxiv},
  pages={2025--05},
  year={2025},
  publisher={Cold Spring Harbor Laboratory}
}

@article{cao2024impact,
  title={The Impact of Model Misspecification on Phylogenetic Network Inference},
  author={Cao, Zhen and Li, Meng and Ogilvie, Huw and Nakhleh, Luay},
  journal={Bulletin of the Society of Systematic Biologists},
  volume={3},
  number={1},
  year={2024},
  publisher={The Ohio State University Libraries}
}

@inproceedings{yakici2022phylogenetic,
  title={Phylogenetic Network Dissimilarity Measures That Take Branch Lengths Into Account},
  author={Yakici, Berk A and Ogilvie, Huw A and Nakhleh, Luay},
  booktitle={RECOMB International Workshop on Comparative Genomics},
  pages={86--102},
  year={2022},
  organization={Springer}
}

@article{cao2019practical,
  title={Practical aspects of phylogenetic network analysis using PhyloNet},
  author={Cao, Zhen and Liu, Xinhao and Ogilvie, Huw A and Yan, Zhi and Nakhleh, Luay},
  journal={BioRxiv},
  pages={746362},
  year={2019},
  publisher={Cold Spring Harbor Laboratory}
}

@Inbook{Elworth2019,
author="Elworth, R. A. Leo
and Ogilvie, Huw A.
and Zhu, Jiafan
and Nakhleh, Luay",
editor="Warnow, Tandy",
title="Advances in Computational Methods for Phylogenetic Networks in the Presence of Hybridization",
bookTitle="Bioinformatics and Phylogenetics: Seminal Contributions of Bernard Moret",
year="2019",
publisher="Springer International Publishing",
address="Cham",
pages="317--360",
}

@incollection{nakhleh2010evolutionary,
  title={Evolutionary phylogenetic networks: models and issues},
  author={Nakhleh, Luay},
  booktitle={Problem solving handbook in computational biology and bioinformatics},
  pages={125--158},
  year={2010},
  publisher={Springer}
}

@article{nakhleh2009metric,
  title={A metric on the space of reduced phylogenetic networks},
  author={Nakhleh, Luay},
  journal={IEEE/ACM Transactions on Computational Biology and Bioinformatics},
  volume={7},
  number={2},
  pages={218--222},
  year={2009},
  publisher={IEEE}
}

@article{bastide2018phylogenetic,
  title={Phylogenetic comparative methods on phylogenetic networks with reticulations},
  author={Bastide, Paul and Sol{\'\i}s-Lemus, Claudia and Kriebel, Ricardo and William Sparks, K and An{\'e}, C{\'e}cile},
  journal={Systematic biology},
  volume={67},
  number={5},
  pages={800--820},
  year={2018},
  publisher={Oxford University Press}
}

@article{jhwueng2015trait,
  title={Trait evolution on phylogenetic networks},
  author={Jhwueng, Dwueng-Chwuan and O’Meara, Brian C},
  journal={BioRxiv},
  pages={023986},
  year={2015},
  publisher={Cold Spring Harbor Laboratory}
}

@article{teo2023accounting,
  title={Accounting for intraspecific variation in continuous trait evolution on a reticulate phylogeny},
  author={Teo, Benjamin and Rose, Jeffrey and Bastide, Paul and An{\'e}, C{\'e}cile},
  journal={Bulletin of the Society of Systematic Biologists},
  volume={2},
  number={3},
  pages={1--29},
  year={2023}
}

@article{teo2025leveraging,
  title={Leveraging graphical model techniques to study evolution on phylogenetic networks},
  author={Teo, Benjamin and Bastide, Paul and An{\'e}, C{\'e}cile},
  journal={Philosophical Transactions B},
  volume={380},
  number={1919},
  pages={20230310},
  year={2025},
  publisher={The Royal Society}
}

@article{karimi2020reticulate,
  title={Reticulate evolution helps explain apparent homoplasy in floral biology and pollination in baobabs (Adansonia; Bombacoideae; Malvaceae)},
  author={Karimi, Nisa and Grover, Corrinne E and Gallagher, Joseph P and Wendel, Jonathan F and An{\'e}, C{\'e}cile and Baum, David A},
  journal={Systematic Biology},
  volume={69},
  number={3},
  pages={462--478},
  year={2020},
  publisher={Oxford University Press}
}

@article{lipson2020applying,
  title={Applying f4-statistics and admixture graphs: Theory and examples},
  author={Lipson, Mark},
  journal={Molecular Ecology Resources},
  volume={20},
  number={6},
  pages={1658--1667},
  year={2020},
  publisher={Wiley Online Library}
}

@article{peter2016admixture,
  title={Admixture, population structure, and F-statistics},
  author={Peter, Benjamin M},
  journal={Genetics},
  volume={202},
  number={4},
  pages={1485--1501},
  year={2016},
  publisher={Oxford University Press}
}

@article{lipson2013efficient,
  title={Efficient moment-based inference of admixture parameters and sources of gene flow},
  author={Lipson, Mark and Loh, Po-Ru and Levin, Alex and Reich, David and Patterson, Nick and Berger, Bonnie},
  journal={Molecular biology and evolution},
  volume={30},
  number={8},
  pages={1788--1802},
  year={2013},
  publisher={Oxford University Press}
}

@article{loh2013inferring,
  title={Inferring admixture histories of human populations using linkage disequilibrium},
  author={Loh, Po-Ru and Lipson, Mark and Patterson, Nick and Moorjani, Priya and Pickrell, Joseph K and Reich, David and Berger, Bonnie},
  journal={Genetics},
  volume={193},
  number={4},
  pages={1233--1254},
  year={2013},
  publisher={Oxford University Press}
}

@article{deng2021distribution,
  title={The distribution of waiting distances in ancestral recombination graphs},
  author={Deng, Yun and Song, Yun S and Nielsen, Rasmus},
  journal={Theoretical Population Biology},
  volume={141},
  pages={34--43},
  year={2021},
  publisher={Elsevier}
}

@article{bloomquist2010unifying,
  title={Unifying vertical and nonvertical evolution: a stochastic ARG-based framework},
  author={Bloomquist, Erik W and Suchard, Marc A},
  journal={Systematic biology},
  volume={59},
  number={1},
  pages={27--41},
  year={2010},
  publisher={Oxford University Press}
}

@article{mostowy2017efficient,
  title={Efficient inference of recent and ancestral recombination within bacterial populations},
  author={Mostowy, Rafal and Croucher, Nicholas J and Andam, Cheryl P and Corander, Jukka and Hanage, William P and Marttinen, Pekka},
  journal={Molecular biology and evolution},
  volume={34},
  number={5},
  pages={1167--1182},
  year={2017},
  publisher={Oxford University Press}
}

@article{vaughan2017inferring,
  title={Inferring ancestral recombination graphs from bacterial genomic data},
  author={Vaughan, Timothy G and Welch, David and Drummond, Alexei J and Biggs, Patrick J and George, Tessy and French, Nigel P},
  journal={Genetics},
  volume={205},
  number={2},
  pages={857--870},
  year={2017},
  publisher={Oxford University Press}
}

@article{barrat2022treeknit,
  title={TreeKnit: Inferring ancestral reassortment graphs of influenza viruses},
  author={Barrat-Charlaix, Pierre and Vaughan, Timothy G and Neher, Richard A},
  journal={PLoS Computational Biology},
  volume={18},
  number={8},
  pages={e1010394},
  year={2022},
  publisher={Public Library of Science San Francisco, CA USA}
}

@article{muller2020bayesian,
  title={Bayesian inference of reassortment networks reveals fitness benefits of reassortment in human influenza viruses},
  author={M{\"u}ller, Nicola F and Stolz, Ugn{\.e} and Dudas, Gytis and Stadler, Tanja and Vaughan, Timothy G},
  journal={Proceedings of the National Academy of Sciences},
  volume={117},
  number={29},
  pages={17104--17111},
  year={2020},
  publisher={National Academy of Sciences}
}

@inproceedings{yurovsky2010flurf,
  title={Flurf, an automated flu virus reassortment finder based on phylogenetic trees},
  author={Yurovsky, Alisa and Moret, Bernard ME},
  booktitle={2010 IEEE International Conference on Bioinformatics and Biomedicine (BIBM)},
  pages={579--584},
  year={2010},
  organization={IEEE}
}

@article{yc2022evaluation,
  title={Evaluation of methods for estimating coalescence times using ancestral recombination graphs},
  author={YC Brandt, D{\'e}bora and Wei, Xinzhu and Deng, Yun and Vaughn, Andrew H and Nielsen, Rasmus},
  journal={Genetics},
  volume={221},
  number={1},
  pages={iyac044},
  year={2022},
  publisher={Oxford University Press}
}

@article{stolz2022joint,
  title={Joint inference of migration and reassortment patterns for viruses with segmented genomes},
  author={Stolz, Ugn{\.e} and Stadler, Tanja and M{\"u}ller, Nicola F and Vaughan, Timothy G},
  journal={Molecular biology and evolution},
  volume={39},
  number={1},
  pages={msab342},
  year={2022},
  publisher={Oxford University Press}
}

@article{muller2022bayesian,
  title={A Bayesian approach to infer recombination patterns in coronaviruses},
  author={M{\"u}ller, Nicola F and Kistler, Kathryn E and Bedford, Trevor},
  journal={Nature communications},
  volume={13},
  number={1},
  pages={4186},
  year={2022},
  publisher={Nature Publishing Group UK London}
}

@misc{phyloplots,
  author = {Cecile An\'e and others},
  title = {https://github.com/JuliaPhylo/PhyloPlots.jl},
  year = 2018,
  url = {https://github.com/JuliaPhylo/PhyloPlots.jl},
  urldate = {2025-07-27}
}

@misc{phylotraits,
  author = {Cecile An\'e and others},
  title = {https://github.com/JuliaPhylo/PhyloTraits.jl},
  year = 2025,
  url = {https://github.com/JuliaPhylo/PhyloTraits.jl},
  urldate = {2025-07-27}
}

@misc{tanggle,
  author = {Klaus Schliep and others},
  title = {BioConductor: tanggle},
  year = 2025,
  url = {https://www.bioconductor.org/packages/release/bioc/vignettes/tanggle/inst/doc/tanggle_vignette.html},
  urldate = {2025-07-27}
}

@article{specht2025ggret,
  title={ggret: An R package for visualising and manipulating tree-based phylogenetic networks},
  author={Specht, Gerd and Schmid, Clemens and K{\"u}hnert, Denise and Kocher, Arthur},
  journal={Journal of Open Source Software},
  volume={10},
  number={110},
  pages={7773},
  year={2025}
}

@book{semple2003phylogenetics,
  title={Phylogenetics},
  author={Semple, Charles and Steel, Mike and others},
  volume={24},
  year={2003},
  publisher={Oxford University Press on Demand}
}

\end{document}